\documentclass[preprint,number,times,sort&compress]{elsarticle}

\usepackage{graphicx}
\usepackage{dcolumn}
\usepackage{bm}
\usepackage{pifont}
\usepackage{natbib}
\usepackage{fleqn}

\usepackage[
]{geometry}

\usepackage{epstopdf}
\begin{document}

\title{Secondary phase Cu$_2$SnSe$_3$ vs. kesterite Cu$_2$ZnSnSe$_4$:\\ 
similarities and differences in lattice vibration modes}
\author{N. B. Mortazavi Amiri}
\author{A. V. Postnikov}
\ead{andrei.postnikov@univ-lorraine.fr}
\address{LCP-A2MC, Institute Jean Barriol,
Universit\'e de Lorraine,
1 Bd Arago, F-57078 Metz, France}
\date{\today}
\begin{abstract}
The crystal structure of monoclinic semiconductor Cu$_2$SnSe$_3$
is optimized, in a first-principles LDA calculation by 
{\sc Siesta} method, to be found in good agreement with 
available experimental data, on which base
zone-center transversal phonon modes
are further calculated. The comparison with a similar calculation
for kesterite-phase Cu$_2$ZnSnSe$_4$ helps to identify 
vibration modes promising to serve as fingerprints for
discrimination of these two materials from their
lattice-dynamical properties. Moreover, a full analysis of
vibration modes is done, which emphasizes an importance of
structural motives present in Cu$_2$SnSe$_3$ but absent in kesterite,
namely continuous planar chains and stripes of like cations/anions,
for the manifestation of structure-specific vibration lines. 
\end{abstract}
\maketitle

\section{Introduction}
In increasingly numerous studies being done 
on quaternary semiconductor systems
promising for photovoltaic applications, kesterite-type
Cu$_2$ZnSnX$_4$ (X = S, Se) \cite{PhTRSA369-1840,AdvMat22-E156,APL94-041903},
a ``secondary phase'' Cu$_2$SnX$_3$
gains importance \cite{JPD43-215403} as a not always welcome, but annoyingly robust,
member of the quaternary phase diagram \cite{TSF519-7513,TSF519-7394}. 
It would be good to reliably identify, and separate, it from other phases, 
as this contamination may deteriorate the photovoltaic properties of
synthesized materials. The difficulty is that the secondary phase
(Cu$_2$SnSe$_3$, or CTSe for the following) has the same underlying
zinc-blende structure (when neglecting the difference between cations)
as kesterite or stannite phases of Cu$_2$ZnSnSe$_4$ (CZTSe),
hence the X-ray diffraction identification is very problematic.
Fortunately, the corresponding vibration spectra 
seem to show pronounced differences.
The difficulty here is that good benchmarks which would help
to pinpoint one or another phase are not yet established as such,
exactly because of the difficulty to prepare, and characterize,
them in a well controlled way. It should be noted that much activity
in preparation of photovoltaically interesting samples occurs in
thin films and not in massive bulk, so that typical problems of
growth and quality control do add to the difficulty of keeping
the CTSe and CZTSe phases apart.

In this perspective, we hope that first-principles calculations,
relying on unambiguous structure information, may help to pinpoint
such benchmarks in phonon spectra. However, our interest is of
more than purely applied character. Given the overall closeness
in chemistry and a tricky structural relation between
the two phases, we can trace how the features of different 
spatial confinement emerge and give rise to different vibration
motives. This may be didactically interesting, pay homage to
``percolation'' view on the vibration modes in semiconductor 
alloys \cite{PRB77-125208,PRB71-115206}
(CTSe being a nice non-trivial natural benchmark 
of an ordered alloy), and have impact on manipulating semiconductor
superstructures.

The paper is organized as follows.
Section \ref{sec:structure} briefly outlines structural relations 
between kesterite
and the secondary phase, more details of which are put into
the Appendix.
Section \ref{sec:calcul} refers to the technical side of calculations.
Section \ref{sec:optimi} specifies the optimizes crystal structure
in comparison with experiment. The main section \ref{sec:modes}
present the discussion of vibration features mode by mode, 
making reference to their counterparts in kesterite
and -- whenever posible -- to experiments, suggesting tentative
attribution of the most pronounced measured modes.

\section{Structural relation between phases}
\label{sec:structure}
The crystal structure of CTSe, 
as reported by Marcano \emph{et al.} \cite{JAP90-1847,MatLet53-151} 
and refined by Delgado \emph{et al.} \cite{MatResBul38-1949},
is monoclinic with the $Cc$ space group.
In fact its unit cell is a supercell of underlying zincblende structure, where two atom species occupy cationic sites in an ordered way
(Cu being in two inequivalent positions),
with $Z$=4, hence 12 cationic sites per unit cell. 
Kesterite is a different ordered superstructure
on the basis of zincblende, with three different cations and
8 cationic sites. The two structures are commensurate, so that
doubled unit cell of CTSe matches tripled unit cell of kesterite 
(not stannite!) CZTSe.

The orientation of the monoclinic structure of CTSe with respect
to underlying cubic is shown in Fig.~1 of Ref.~\cite{MatLet53-151}.
Two vectors of the monoclinic basal plane, $\mathbf{a}$ and $\mathbf{c}$,
go as $(\frac{1}{2}\;\frac{1}{2}\;{\pm}1)$ and are thus valid translation
vectors of kesterite. The $\mathbf{b}$ vector is in the basal plane
of kesterite, $(\frac{3}{2}\,-\!\frac{3}{2}\;0)$, which, 
if starting from a Cu atom in the Cu-Sn plane, would end at Sn. 
Therefore this vector
must be doubled in order to get a structure comprising an integer
number (3) of kesterite unit cells. More precisely the transformation
of basis vectors is described in Appendix. 
We only emphasize here, as was also stated 
by Delgado \emph{et al.},\cite{MatResBul38-1949} that copper atoms
have two inequivalent positions, whereby selenium has three.
The nearest neighborhoods of anions are 
(Cu1, Cu2, 2$\times$Sn) to Se1,
(Cu1, 2$\times$Cu2, Sn) to Se2, 
(2$\times$Cu1, Cu2, Sn) to Se3
and (of cations)
(Se1, Se2, 2$\times$Se3) to Cu1,
(Se1, 2$\times$Se2, Se3) to Cu2,
(2$\times$Se1, Se2, Se3) to Sn.
The explicit numbering and positioning of atoms is given
in the Appendix.

\section{Calculation aspects}
\label{sec:calcul}
The present calculations have been done from first principles,
at the accuracy level provided by the density functional theory,
specifically, the local density approximation (LDA),
using the {\sc Siesta} method and calculation 
code \cite{siesta,PRB53-R10441,JPCM14-2745}. Norm-conserving
pseudopotentials constructed with the Troullier-Martins 
scheme \cite{PRB43-1993} were used in combination with
atom-centered strictly confined localized basis functions.
The calculation setup is identical to that used 
in Ref.~\cite{PRB82-205204}. The calculations on the kesterite phase
have not been re-done here, as compared to this publication, 
but we bring to attention some further vibration modes 
than previously discussed, for the sake of the present comparison.

An obvious shortcoming of our calculation is the underestimation
of the band gap and, as a consequence, a very limited usefulness
for discussing electronic properties. However, it has been shown
on many occasions that ground state-related results, such as
equilibrium geometry, forces on atoms and hence phonons, are
quite reliable, so that more sophisticated hybrid-functional
or GW calculations are not (yet) prerequisite of a reliable
lattice-dynamics study.
As concerns the electronic structure with the accent on optical
properties, the necessary information was provided by a recent
work by Ying-Teng Zhai \emph{et al.} \cite{PRB84-075213}, 
who did a hybrid-functional calculation for monoclinic CTSe,
among other ternary structures.

Another limitation, that we only discuss the zone-center vibration
modes, is of purely technical character and could have been
overcome in our calculation approach. However, for discussing
a possible impact on Raman or infrared spectra, only zone-center
vibrations are of interest anyway. Moreover, having already 
a quite large unit cell of the CTSe phase and hence 
correspondingly reduced Brillouin zone, discussing dispersions 
of its many phonon branches would not obviously contribute
any clearness to the present analysis. And finally, having different
unit cells of CZTSe-kesterite and CTSe, an attribution of their
relative phonon dispersions would be not straightforward. 
At $\Gamma$, such comparison is easier.

\section{Structure optimization}
\label{sec:optimi}
The internal coordinates of atoms in monoclinic CTSe
were given by Delgado \emph{et al.}.\cite{MatResBul38-1949}
The details on the crystal structure and its relation to kesterite
are explained in the Appendix; for the present it suffices to know
that the underlying lattice is a very little distorted zincblende,
over cationic sites of which the Cu and Sn atoms are ordered. 

In the course of relaxation, even as no symmetry constraints 
have been imposed on the lattice, the latter remained well 
monoclinic [$\alpha$ and $\gamma$ stay at (90$\pm$10$^{-4}$)$^{\circ}$],
yielding lattice parameters $a$=6.939~{\AA}, $b$=11.950~{\AA},
$c$=6.975~{\AA}, $\beta$=109.72$^{\circ}$ and unit cell volume
$V$=544.45~{\AA}$^3$. These can be compared with 
experimental data of Ref.~\cite{MatResBul38-1949},
$a$=6.967~{\AA}, $b$=12.049~{\AA}, $c$=6.945~{\AA}, 
$\beta$=109.19$^{\circ}$, $V$=550.6~{\AA}$^3$
($a$ and $c$ apparently being inversed, compared to our setting).
The underestimation of lattice parameters in LDA is typical;
the present error (1\%, for the volume)
is in fact unusually small. We note however that earlier experimental
data by Marcano \emph{et al.} \cite{JAP90-1847,MatLet53-151}
($a$=6.5936~{\AA}, $b$=12.1593~{\AA}, $c$=6.6084~{\AA}, $\beta$=108.56$^{\circ}$) 
are rather at variance with both Delgado's results and our present ones.

An attribution of different sites with their respective coordinates,
as extracted from X-ray diffraction, is given in Table~3
of Ref.~\cite{MatResBul38-1949}. As the {\sc Siesta} calculation
does not impose symmetry constraints, the coordinates of all 24 atoms
are independently adjusted, along with lattice parameters, in the course
of relaxation. In order to facilitate the comparison with experiment,
the nominal fractional coordinates within each group of (four) atoms
are averaged, upon applying symmetry transformations between corresponding
Wyckoff positions; the results are given in Table~\ref{tab:struc1}.

\begin{table}
\caption{\label{tab:struc1}
Internal coordinates in CTSe as presently calculated
and reported in Ref.~\cite{MatResBul38-1949}
(all in $4a$ Wyckoff positions of the $Cc$ space group).
The calculated values are averages over four
nominally equivalent positions within each type; see text for
details.
}
\begin{center}
\begin{tabular}{lr@{.}lr@{.}lr@{.}l}
\hline
Type & \multicolumn{2}{c}{$x$} & \multicolumn{2}{c}{$y$} & \multicolumn{2}{c}{$z$} \\  
\hline
    & \multicolumn{6}{c}{Calculation} \\
Cu1 &    0&3864 & 0&2535 &    0&6206 \\
Cu2 &    0&3983 & 0&4175 &    0&1206 \\
Sn  &    0&3772 & 0&0914 &    0&1015 \\
Se1 &    0&0344 & 0&4023 &    0&0090 \\
Se2 & $-$0&0207 & 0&0838 & $-$0&0209 \\
Se3 &    0&5261 & 0&2660 & $-$0&0203 \\
\hline
    & \multicolumn{6}{c}{Experiment} \\
Cu1 &    0&371(3) & 0&257(1) &    0&616(3) \\
Cu2 &    0&370(3) & 0&418(1) &    0&116(3) \\
Sn  &    0&363(3) & 0&091(1) &    0&107(3) \\
Se1 &    0&000    & 0&409(1) &    0&000    \\
Se2 & $-$0&026(3) & 0&078(1) & $-$0&015(3) \\
Se3 &    0&503(3) & 0&259(1) & $-$0&014(3) \\
\hline
\end{tabular}
\end{center}
\end{table}    

\begin{table}
\caption{\label{tab:struc2}
Bond lengths in CTSe according to present calculation
and X-ray diffraction data of Ref.~\cite{MatResBul38-1949}.
}
\begin{center}
\begin{tabular}{lccc}
\hline
 & Cu1 & Cu2 & Sn \\
\hline
 & \multicolumn{3}{c}{Calculation} \\
Se1 & 2.380 & 2.384 & 2.679 \\
Se2 & 2.366 & 2.370 & 2.602 \\
Se3 & 2.367 & 2.370 & 2.595 \\
\hline
 & \multicolumn{3}{c}{Experiment} \\
Se1 & 2.44(2) & 2.44(2) & 2.59(2) \\
Se2 & 2.39(2) & 2.38(7) & 2.57(3) \\
Se3 & 2.43(3) & 2.43(3) & 2.51(2) \\
\hline
\end{tabular}
\end{center}
\end{table}    

\begin{figure}
\centerline{\includegraphics[width=0.75\textwidth]{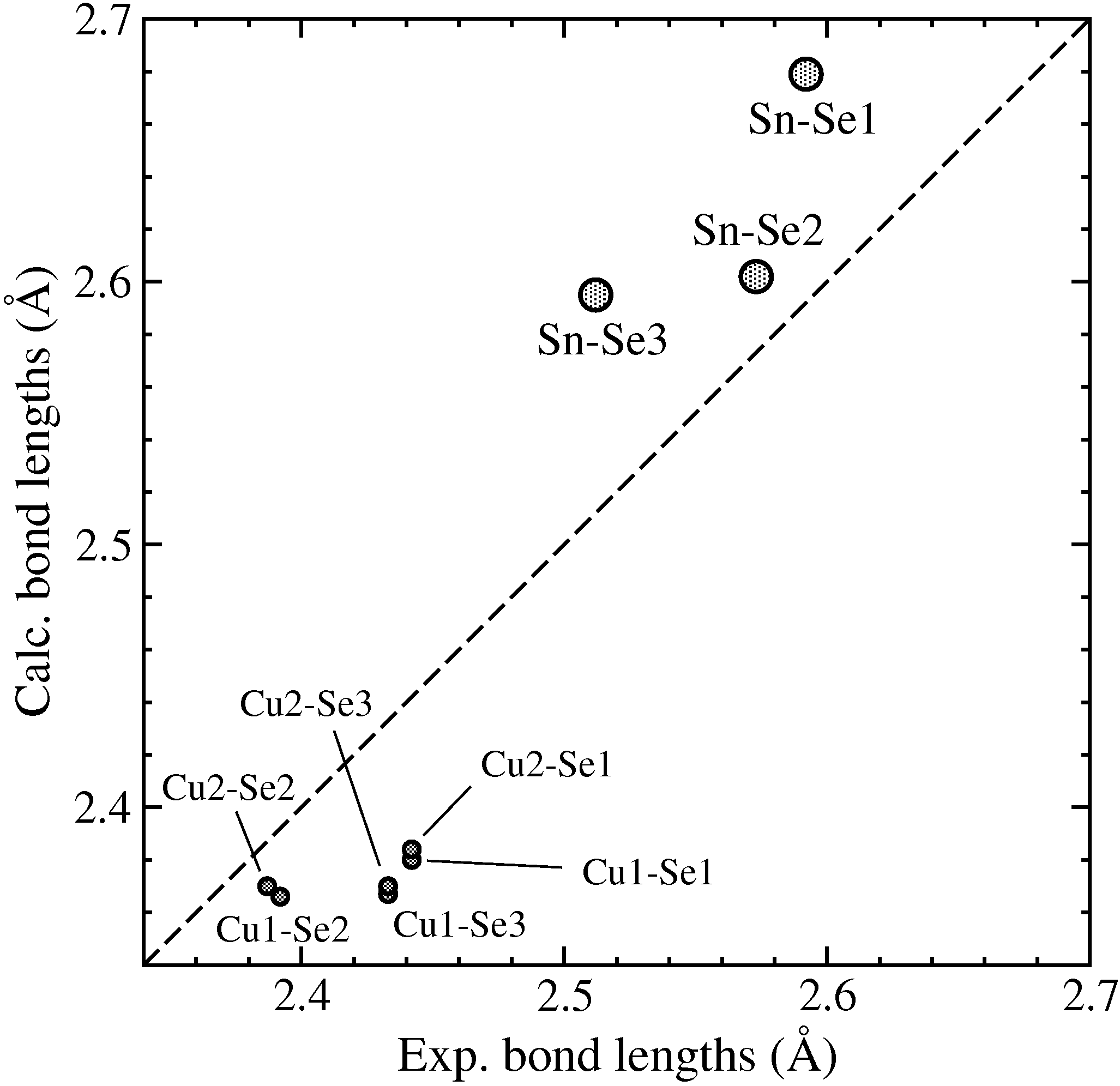}}
\caption{\label{fig:bonds}
First-neighbor interatomic distances
from calculated structure relaxation
in comparison with experimental data 
of Ref.~\cite{MatResBul38-1949}.  
}
\end{figure}

The resulting bond lengths are summarized in Fig.~\ref{fig:bonds}.
On top of fairly good overall agreement one can note that 
Sn--Se bonds are systematically overestimated in calculation,
at the expense of reduced Cu--Se ones. Put differently,
the Se1 and Se2 atoms are off-centered in their respective
cation tetrahedra more versus Cu and away from Sn than they should. 

\section{Discussion on phonons}
\label{sec:modes}
The zone-center frozen phonon calculation on CTSe yields 
3 (acoustic, zero-frequency) + 69 further modes, which, 
atomic differences within the each sublattice neglected, would map
a number of optical-like and acoustical-like vibrations
at some $\mathbf{q}$ values of the underlying
zincblende lattice. Of primary practical interest is
a comparison with Raman and infrared spectra, probing
$\mathbf{q}$=0 vibrations. Unable to strictly calculate
corresponding intensities, we use the technique of projection
of phonon eigenvectors onto a vibration pattern corresponding
to a given $\mathbf{q}$
(see Eq.~1 of Ref.~\cite{PRB82-205204}),
notably $\mathbf{q}$=0 in the present case.
This strictly suppresses vibration which involve atoms of the same
species in counterphase. The rest of analysis, e.g., identification
of ``optical-like'' modes in which cation vibrate against
anions rather than in phase, has to be done by immediate inspection.                          

\begin{figure}
\centerline{\includegraphics[width=0.82\textwidth]{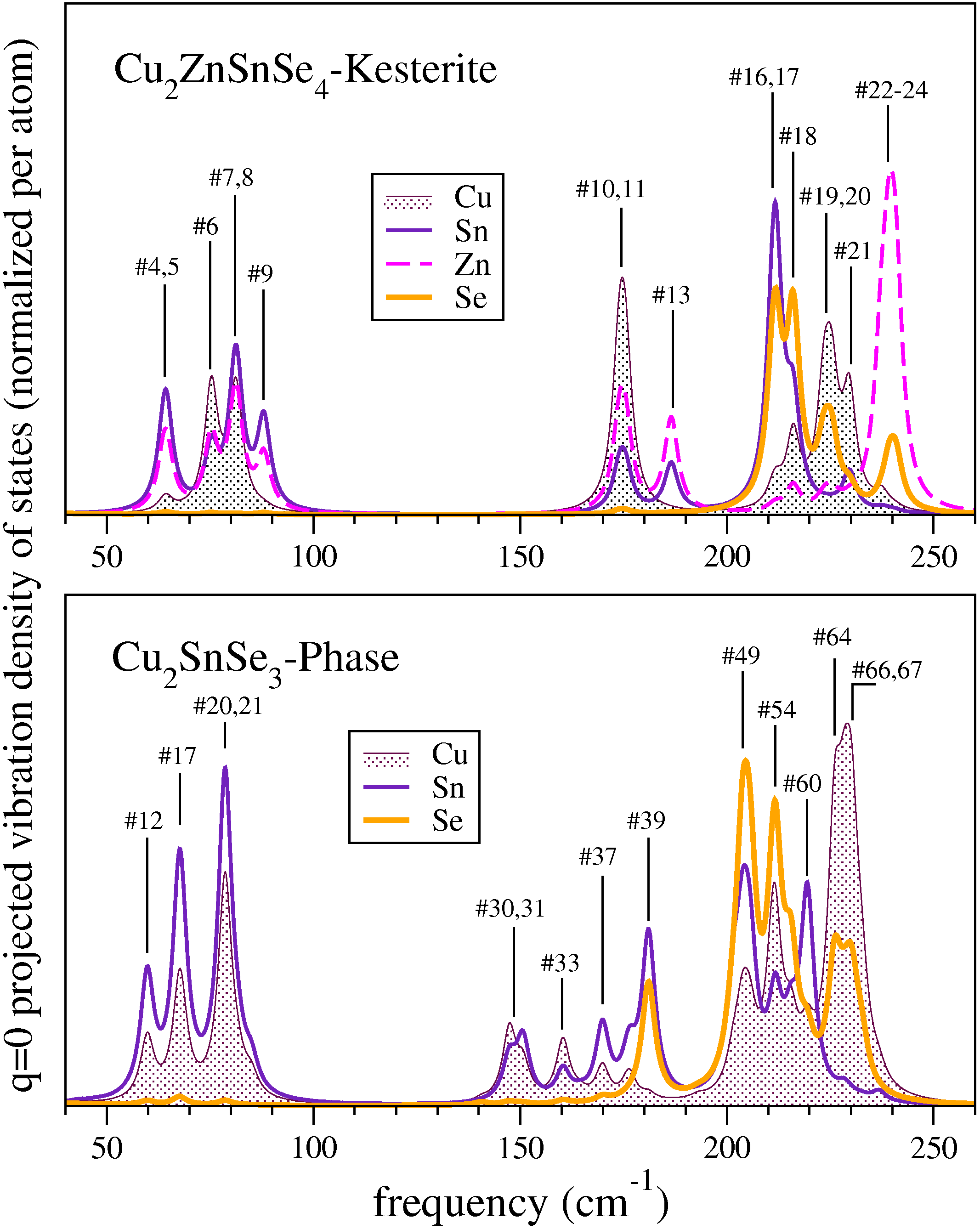}}
\caption{\label{fig:phdos1}
(color online)
Calculated zone-center density of modes in CZTSe-kesterite
(top panel, from Ref.~\cite{PRB82-205204})
and CTSe (bottom panel, present calculation). Some modes in throughout
numbering are indicated, for reference in the text.
}
\end{figure}

Fig.~\ref{fig:phdos1} compares the vibration density of states,
calculated along Eq.~(1) of Ref.~\cite{PRB82-205204}
with $\mathbf{q}=0$, for kesterite structure of CZTSe and
for CTSe. The upper panel 
in fact depicts the system
of modes earlier published in Fig.~5 (middle panel) of
Ref.~\cite{PRB82-205204}, but now plotted with better resolution
(broadening halfwidth of 2~cm$^{-1}$). The character of each mode
is explained in Table~I of Ref.~\cite{PRB82-205204}. 

Observing the overall similarity in the placement of three 
separated dense groups of vibration lines
(50-90~cm$^{-1}$, 140-190~cm$^{-1}$ and 200-250~cm$^{-1}$,
that equally applies to CZTSe stannite and CuInSe$_2$
chalcopyrite, see Ref.~\cite{PRB82-205204}),
we note, as the most striking difference, that in CTSe
the second group is broader and gets softened. 
Two other groups roughly maintain their widths as in kesterite, 
exhibiting however a slight red shift, and their composition
(in what concerns the involvement of different atoms in vibrations)
is sometimes different. We note in passing that the counting of modes
within each group is conform between kesterite and CTSe:
having a tripled number of atoms and hence of zone-center modes
in the latter phase,
e.g., the modes 10 through 13 of kesterite are ``replaced'' by modes
30 through 39 in CTSe.   

Given a moderate number of atoms in the primitive cells and
a fair amount of residual symmetry in corresponding phases,
non-negligible symmetric patterns can be found in, at least,
some phonon modes of kesterite (easy to see) and of CTSe
(somehow hidden). The intricacy of such comparison is that symmetry
patterns of tetragonal kesterite (displacements parallel
to edges, or basal diagonal) are replaced to symmetry of
``almost hexagonal'' planes, ``hidden'' in the CTSe structure.
In the following, we briefly discuss the nature of some
vibration modes responsible for the pronounced, or otherwise
interesting, peaks in the zone-center density of modes
shown in Fig.~\ref{fig:phdos1}.

\subsection{The softest modes (50 -- 90~cm$^{-1}$)}
The vibrations within this lower group 
are, basically, zone-boundary acoustic modes of the
zincblende aristotype; their non-zero projection is only due
to the non-equivalence of cations, whereas the contributions
from Se do largely cancel out. In these modes, cations tend to move 
in phase with, at least, some neighboring anions; 
such ``acoustic'' movements of certain groups of atoms 
occur in opposite phase to the other ones. 
In the \textbf{mode~\#12} (59.8~cm$^{-1}$), 
the \mbox{-Cu1-Se3-Cu1-Se3-} chains
transversing the crystal along [101] do rigidly move
parallel to their own direction, against the 
(Cu2+Sn+Se1+Se2) bulk. In the mode~\#13 (60.1~cm$^{-1}$), 
two adjacent (010) planes that make such (Cu2+Sn+Se1+Se2) ``layer''
(see details of structure in Appendix)
undergo rigid (in-plane) movement in the opposite sense, 
whereas the third interlacing (2$\times$Cu1+2$\times$Se3) plane, 
containing the previously mentioned chains, does not move.
As a result, this mode yields zero contribution to the 
$\mathbf{q}$=0 -projected density of modes.
The \textbf{mode \#17} at 67.7~cm$^{-1}$ involves almost perfectly planar
movement, somehow scattered within the (010) planes around
the general [001] direction, of Sn with Se1, roughly, in one sense, 
and of Cu2 with Se2 --
in the opposite one, Cu1 and Se3 being, again, not much involved.

\begin{figure}
\centerline{\includegraphics[width=0.86\textwidth]{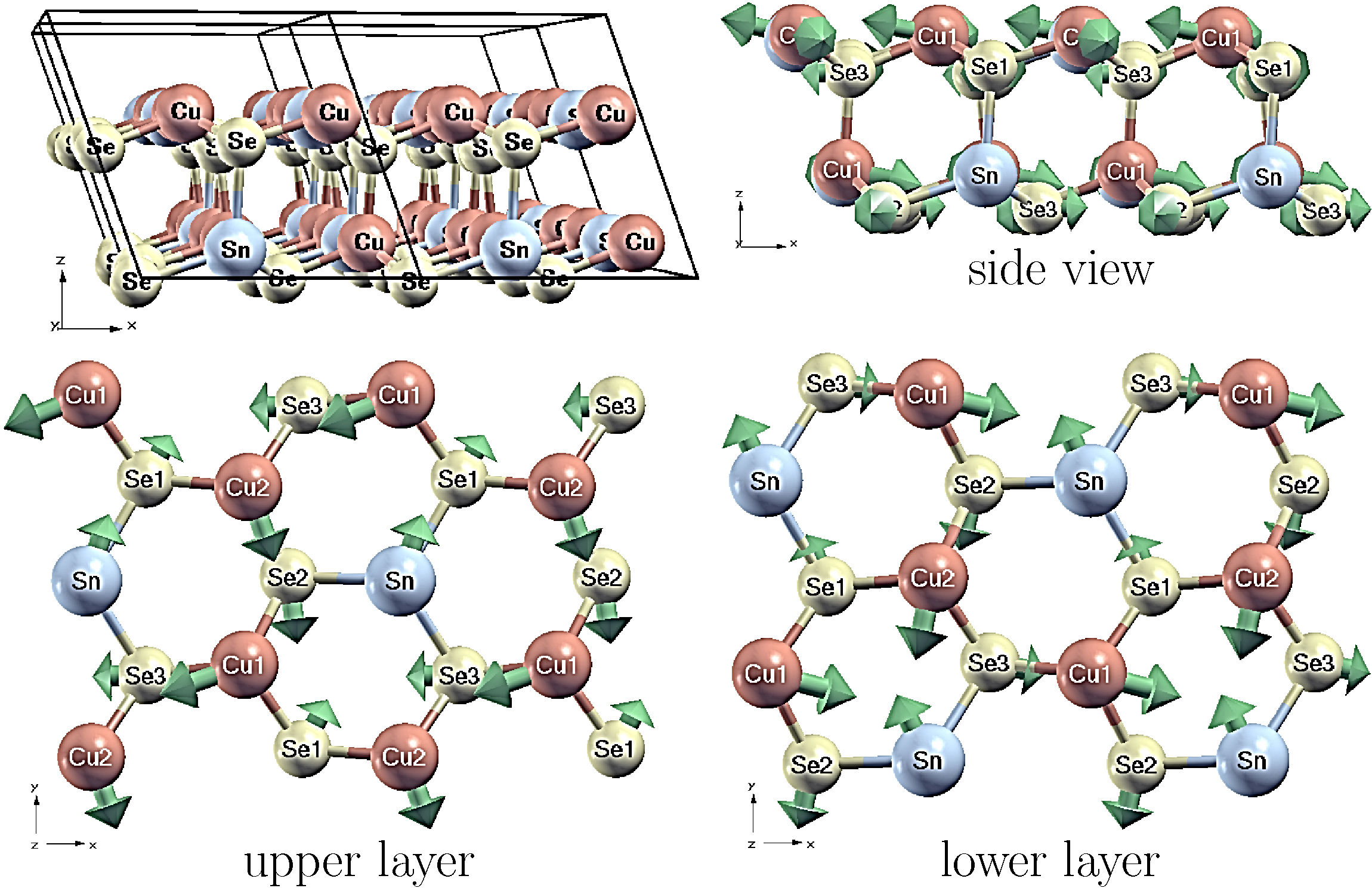}}
\caption{\label{fig:mode20}
(color online)
Snapshot of the vibration mode \#20 at 79~cm$^{-1}$
in two adjacent (001) hexagonal layers 
(Se1 of the upper layer docked on top of Sn in the lower layer)
and in the side view along [010].
The placement of layers in the unit cell is shown in the upper left panel.}
\end{figure}

As frequencies increase within this group of lines,
the folding of the zone-boundary modes becomes less evident;
the nearest cations and anions gradually go out of phase,
developing a ``more optical'' character of modes.
The \textbf{mode \#20} (78.6~cm$^{-1}$) still retains 
the tendencies of neighboring
cations and anions to move together and not produce much
stretching; however, their displacements at almost right angle
are common, and the resulting pattern of bond bendings
becomes quite rich. The vibrations in the mode \#20
are essentially confined to (001) planes
of the monoclinic CTSe structure, shown in the view of the crystal lattice
along [010] in the top left panel in Fig.~\ref{fig:mode20}.
The definition of monoclinic lattice vectors
given in the Appendix makes apparent that such planes are (111) ones
of the underlying zincblende aristotype (the difference
between cations being neglected), piled as warped double-layers 
of cation-anion honeycomb network. Each double-layer maintains
the stoichiometry of the compound, having all six inequivalent atoms
to appear along the perimeter of each hexagon. The presence 
of a tin atom at one vertex distorts somehow the ``hexagonality''
within these planes of the relaxed structure. However, the vibration
pattern within each plane is nicely symmetric: out of three hexagons per
unit cell in the plane, one flips back and forth between two
trigonal distortions,
while the others around it make a roughly rigid rotational movement.
The vibrations within both layers are of course identical,
however with trigonal distortion (cations inwards / anions outwards) 
being in opposite phase, and vibration pattern as a whole
being in-plane displaced and rotated.
The mode \#21, degenerate with the present one, looks identical,
only that, the displacement vector on each individual atom being 
rotated by 90$^{\circ}$, a previously ``deforming'' hexagon
becames a ``rigid rotating'' one, and vice versa.  
The resulting movement of all Se atoms sums up to almost zero
and hence disappears in the density of modes of Fig.~\ref{fig:phdos1},
whereas the resulting movement of Sn over both planes
is large, and points opposite to that of Cu. Even if this looks
like an overall Sn vs. Cu mode, the Cu--Se and Sn--Se bonds 
in the deformed hexagons undergo a strong bending, so that 
the prominence of this mode in the resulting spectrum
might be important. 

From the frequency and composition of modes \#20,21 in CTSe,
their ``natural counterparts'' in the kesterite CZTSe seem to be
the modes \#7,8. However, a careful inspection of the vibration
patters in the latter gives no hint of their confinement 
to the zincblende (111) planes -- or, equivalently, the (112) planes
of the kesterite structure. 
Similarly, no ``hexagon dance'' becomes apparent
in these modes of kesterite, whereas their attribution to
the axes of the tetragonal cell, and the corresponding nature of their 
degeneracy, as was previously discussed in Ref.~\cite{PRB82-205204},
are unambiguously pronounced.

\begin{figure}
\centerline{\includegraphics[width=0.88\textwidth]{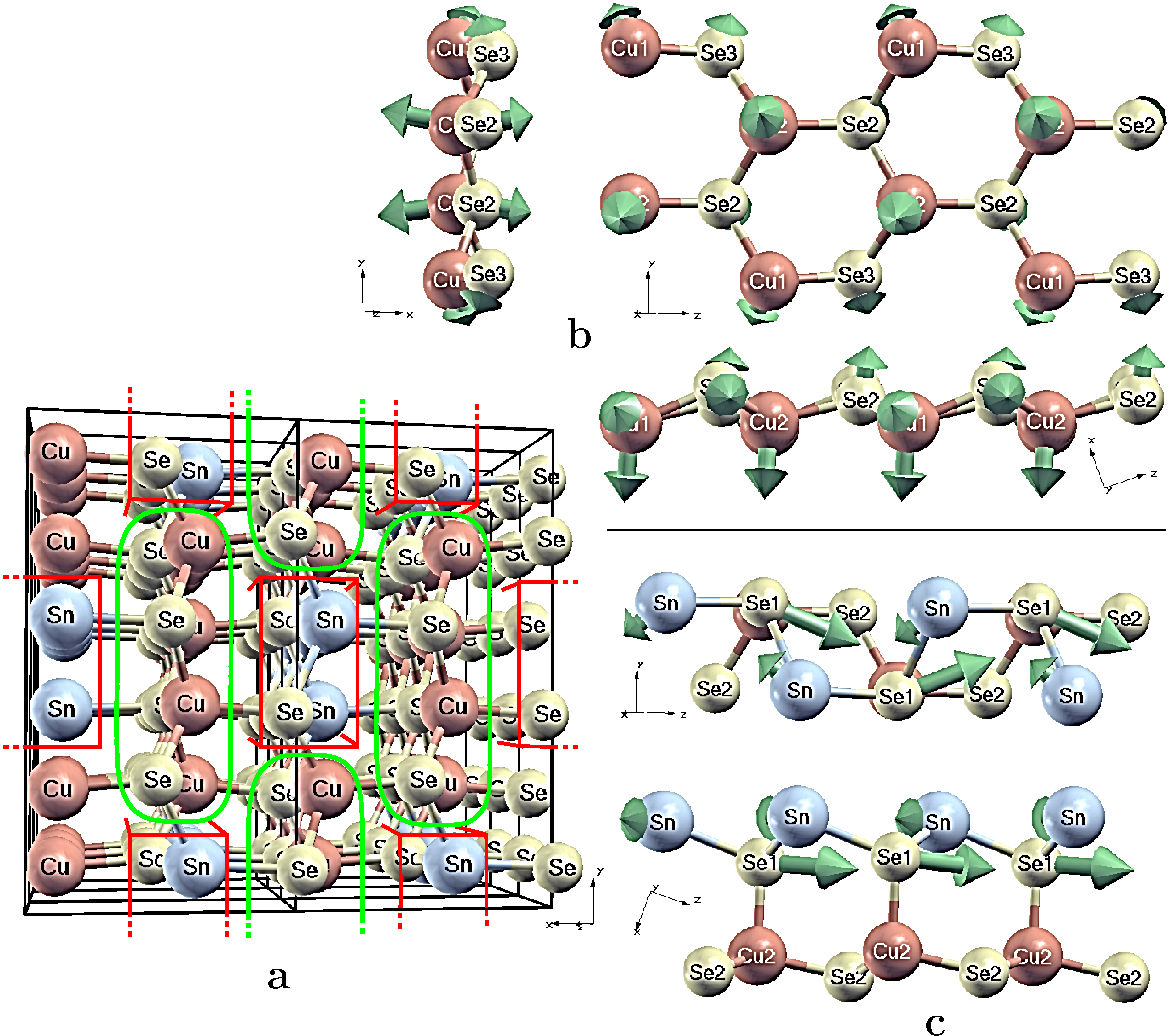}}
\caption{\label{fig:modes30+39}
(color online)
a) Crystal structure of CTSe, viewed along [001].
Isolated -Sn-Se1-Sn-Se1- stripes, along which the mode \#39
propagates (perpendicular to the plane of figure), are outlined
by red boxes. -Cu2-Se2-Cu2-Se2- stripes bordered by 
-Cu1-Se3-Cu1-Se3- chains, running in the same direction,
which are relevant for the mode \#30, are outlined by green ovals.
b) Snapshot of the vibration mode \#30 at 147~cm$^{-1}$ 
with Cu1 and Se2 vibrating perpendicular to ``their'' stripe,
shown in three projections.
c) Snapshot of the vibration mode \#39 at 181~cm$^{-1}$
with Sn and Se1 vibrating along ``their'' stripe, 
shown in two projections.}
\end{figure}

\subsection{Intermediate modes (140 -- 190~cm$^{-1}$)}
Here, we enter the range of ``genuine'' optical modes.
The gradual increase of intensity in the $\mathbf{q}$=0 -projected
density of modes towards the upper end of the spectrum 
reveals that the ``all cations vs. all anions''
character becomes more prominent with frequency. 
A typical motif of vibration 
is cation vs. anion movement at an angle to their connecting bond,
i.e., vibrations are of markedly bond-bending type. Somehow simplifying,
one can note that the displacement-to-bond angle gradually decreases 
as frequency grows throughout the 140 $\rightarrow$ 190~cm$^{-1}$,
so that the upper vibration modes are already quite in the
bond-stretching regime.

The softest in this group, the \textbf{mode \#30} (147.4~cm$^{-1}$)
is confined within the -Cu2-Se2-Cu2-Se2- stripe, a zigzag
chain slightly warped around the (100) plane and running along [001]
of the monoclinic structure. The edges of such stripes are formed by
the -Cu1-Se3-Cu1-Se3- chains, which are knitted together into the
purely Cu1,Se3 (010) planes. The cross-sections of such stripes
are outlined in Fig.~\ref{fig:modes30+39}a by green ovals.
The edge Cu1-Se3 atoms vibrate only weakly, the atoms along each
given [001] line being in phase (see Fig.~\ref{fig:modes30+39}b),
but in alternating sense from one stripe to the neighboring one --
a relict of acoustic zone-boundary behavior. On the contrary, 
the inner Cu2 and Se2 atoms vibrate ``optically'' 
and in ``zone-center way'', 
all Cu2 in phase throughout the crystal against all Se2, that
bends the bonds within the stripe. The mode \#31 (149.0~cm$^{-1}$)
differs in that the sense of Cu2, Se2 vibrations alternates
in consecutive (100) planes; moreover the magnitude
of Se2 displacements reduces considerably, and the stripe-edge atoms,
Cu1 and Se3, vibrate yet weakly but now in opposite to each other.
Due to general inversion of all displacements when passing 
from one (100) plane to the next chemically identical one,
the net zone-center projection of this mode yields zero.
The mode \#32 (150.7~cm$^{-1}$) roughly retains the same system 
of stripes and out-of-plane vibrations, but now the displacements of 
consecutive Cu2 \emph{along the stripe} occur in opposite,
the vibration of in-between Se2 disappears, whereas Cu1 develop
a net (in-phase) component of vibration out of (010) Cu1,Se3-planes.
Se2 and Se3 do not participate much in this mode, whereas the Se1 are
involved in a (largely acoustic -- zone-boundary) interplay with Sn.   
The mode \#33 (160.2~cm$^{-1}$) is difficult to make sense of;
in any case, a transparent system of stripes and either in-phase
or plane-by-plane alternate vibration motives is gone, without
being replaced by any new emerging clear symmetry. 

\begin{figure}
\centerline{\includegraphics[width=0.52\textwidth]{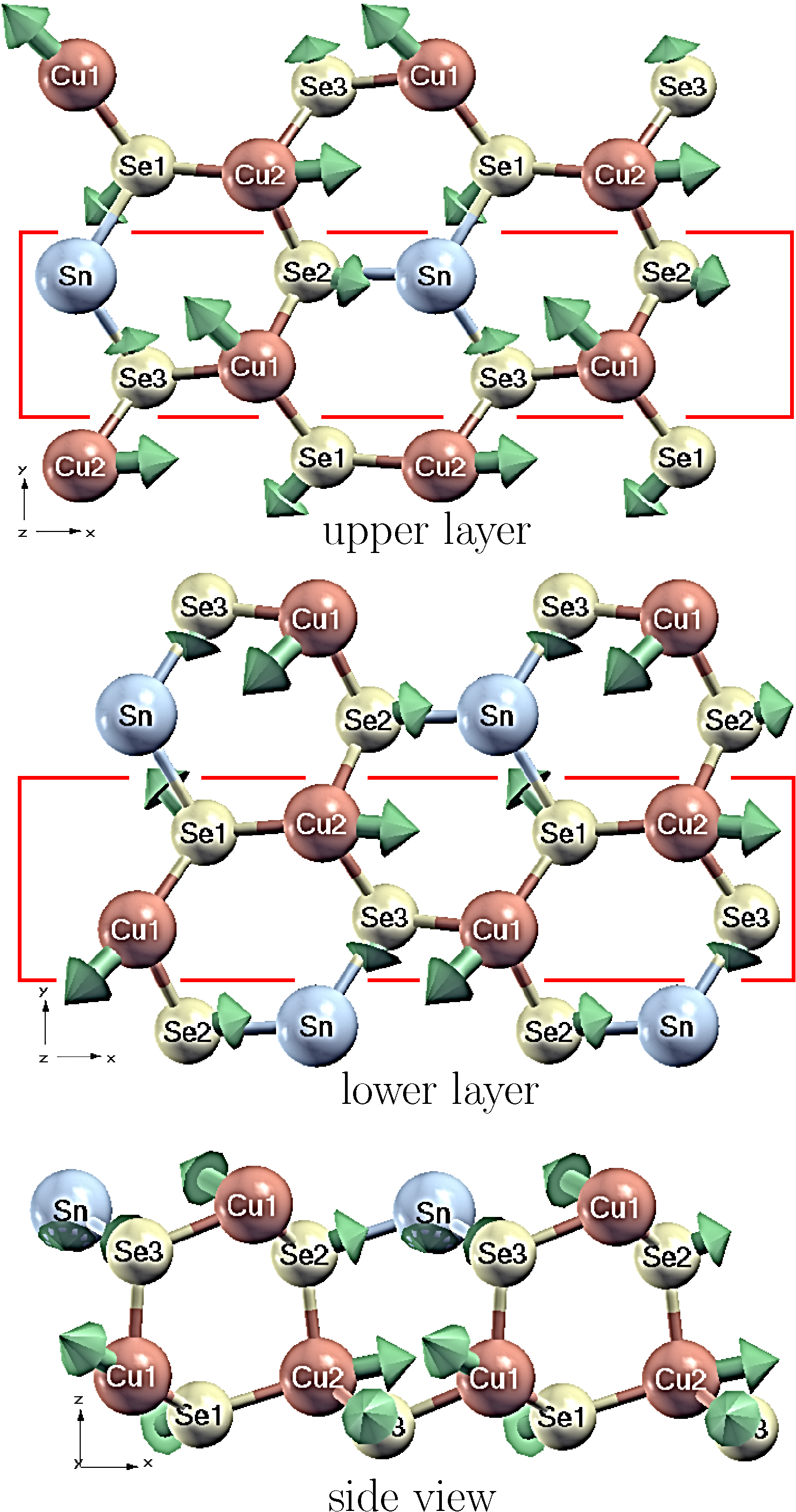}}
\caption{\label{fig:mode37}
(color online)
Snapshot of the vibration mode \#37 at 170~cm$^{-1}$
in two adjacent (001) hexagonal layers and in the side view along
[010], the setting similar to that of Fig.~\ref{fig:mode20}.
The side view shows only atoms within the box
the red trace of which is shown on both layer projections.}
\end{figure}

Such different symmetry reappears in the \textbf{mode \#37}
(169.9~cm$^{-1}$) which is roughly planar 
in the (001) plane, as was mode \#20 (Fig.~\ref{fig:mode20}).
More precisely, the atoms tend to move slightly upwards
or downwards from these planes, along the bonds
to their neighbors in the warped hexagonal layers. A snapshot of this mode
is shown in Fig.~\ref{fig:mode37} in two adjacent planes and
a side view along [010]. 
As previously emphasized for other modes from the middle
(140 -- 190~cm$^{-1}$) group, a vibration of neighboring cations
against anions at some angle, leading to bond bending rather than
clear stretching, remains here a marked element.
However, some elements of stretching are not to overlook:
first of all, a nicely symmetric ``breathing'' 
of three (different) Se anions occurs around an immobile Sn;
moreover, a pair of Cu1 and Cu2 in their concerted motion
stretches their bonds to Se1. 
Taken together, the periodic vibration pattern is such that Cu2
moves roughly parallel to its two neighboring Se2, one being
in the (001) plane and the other in the adjacent plane. 
Tracing their further connections, we can recognize the 
-Cu2-Se2-Cu2-Se2- stripes already discussed above and note that
the Cu2 vibrate perpendicular to the plane of these stripes,
being accompanied by Se2 moving in phase.
However, the vibrations in the rest of crystal are not clear
in this perspective, that's why we don't show the mode \#37 
in the ``stripe'' projection like in Fig.~\ref{fig:modes30+39}.

One notes that Cu1 moves at roughly the right angle 
relative to its Se1 and Se2 neighbors.
Another observation concerning the mode \#37 is that the vibration patterns
in two adjacent (001) planes are not in opposite phase,
as is the case e.g. for the mode \#20 shown in Fig.~\ref{fig:mode20},
but in-phase. More precisely, their vibration patterns are related
by (010) mirror-plane symmetry, as also are the very structures of
adjacent planes -- see Appendix. 

For the discussion of \textbf{mode \#39} (181.1~cm$^{-1}$),
we come back to Fig.~\ref{fig:modes30+39}a
depicting the stripes running along the [001],
but now concentrate on the -Sn-Se1-Sn-Se1- stripes, 
whose cross-sections are outlined by red boxes.
Such stripes are \emph{isolated}, i.e., completely surrounded
by the (Cu1,Cu2,Se2,Se3) matrix which is out of resonance in this mode. 
The snapshot of this vibration (Fig.~\ref{fig:modes30+39}b) shows
that Sn and Se atoms move in opposite, each one towards its next
homologue along the stripe. 
For the mode \#40 nearly degenerate with the latter (181.6~cm$^{-1}$),
the displacement pattern
within each stripe is almost exactly the same, however adjacent stripes,
separated on the lattice by -Cu-Se-Cu-Se- links, move in counterphase.
The $\mathbf{q}$=0 -projected density, summed up over all Sn atoms,
for this mode is nearly zero, but, considering that the stripes are
distant and rather independent, one can expect a considerable 
Raman signal from this mode as well.

In fact these modes are natural suspects for associating them with
a strong Raman line observed by 
Altosaar \emph{et al.} \cite{PSSA205-167},
Grossberg \emph{et al.} \cite{ThinSF517-2489} and
Marcano \emph{et al.} \cite{SSC151-84}
at 180~cm$^{-1}$ in CTSe. We remind that in both the experiment
and our calculation, the line in question 
markedly falls in between two strong peaks attributed to 
the CZTSe kesterite (modes \#10,11 at 174~cm$^{-1}$ 
and \#13 at 187~cm$^{-1}$ in the upper panel of Fig.~\ref{fig:phdos1},
tentatively attributed in Ref.~\cite{PRB82-205204}
to measured lines at 173 and 196~cm$^{-1}$, e.g.,
in Ref.~\cite{PSSA205-167}).

\begin{figure}
\centerline{\includegraphics[width=0.90\textwidth]{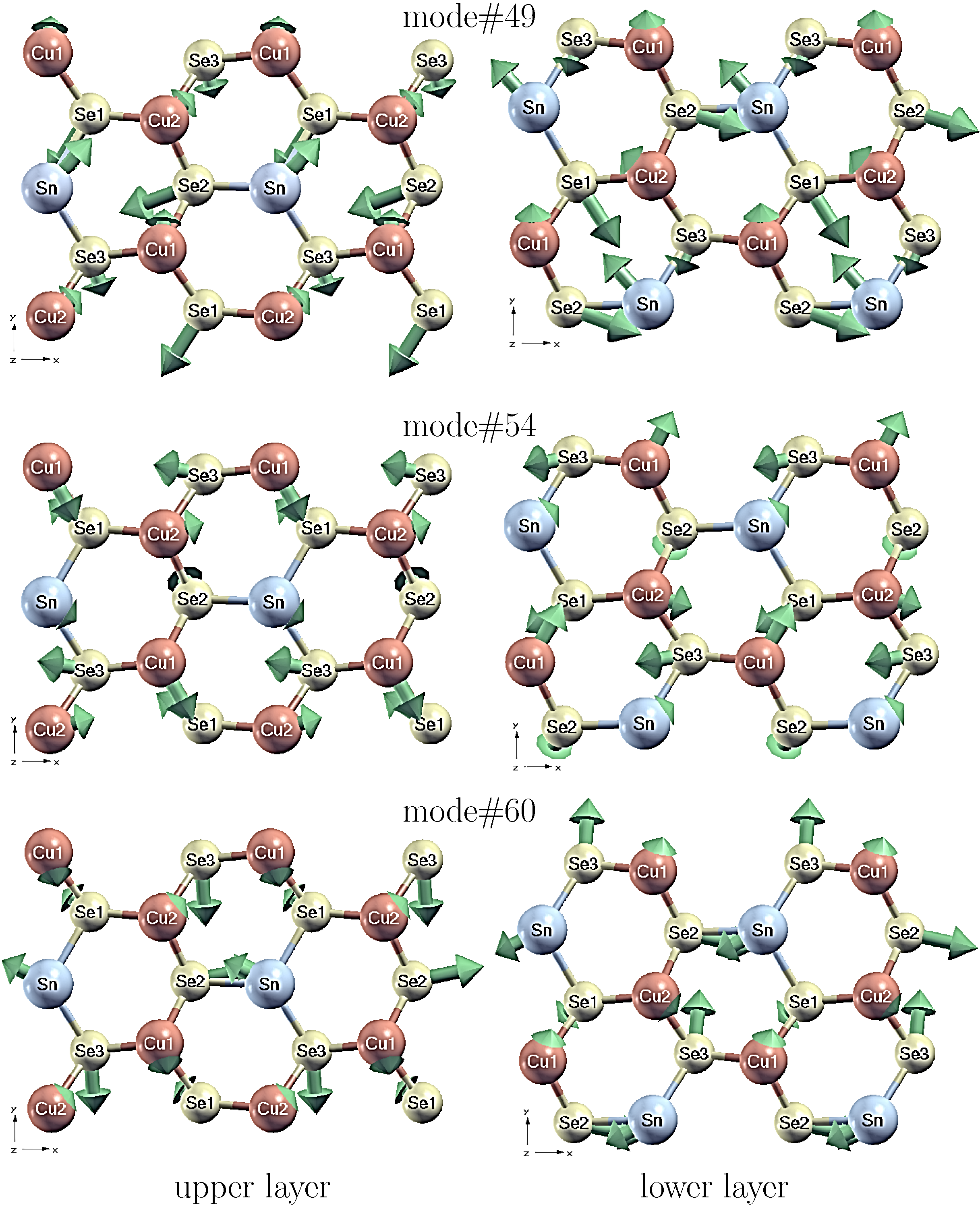}}
\caption{\label{fig:modes49+54+60}
Snapshots of the vibration modes \#49 (at 204~cm$^{-1}$,
upper row), \#54 (at 212~cm$^{-1}$, middle row)
and \#60 (at 219~cm$^{-1}$, bottom row)
depicting the atoms in two consecutive warped (001) planes
(left/right column), similar to those shown in Fig.~\ref{fig:mode20}.
}
\end{figure}

\subsection{The hardest modes (200 -- 240~cm$^{-1}$)}
We turn now to the third group of vibration modes, which
exhibit more pronouncedly bond-stretching character.
Vibrations in the \textbf{mode \#49} (203.8~cm$^{-1}$)
are confined to (001) planes 
of the monoclinic CTSe structure (already discussed above in relation
with the modes~\#20 and \#37) to the extent roughly similar to that
shown in the ``side view'' projection of mode \#20  
in Fig.~\ref{fig:mode20}. The in-plane vibration pattern
in two adjacent planes is shown in the upper row 
of Fig.~\ref{fig:modes49+54+60},
for the in-plane fragments identical to those 
depicted in Fig.~\ref{fig:mode20}; we recall that
Se1 of the upper layer resides on top of Sn from the
lower layer. The key feature of the \#49 vibration mode is
the stretching of the Sn--Se1 bond, which would mark the mirror plane
in an isolated warped layer. 
At the same time, Se2 and Se3, which move
at right angle one to the other, produce a ``resulting motion''
roughly parallel to that of Se1 and, hence, against the Sn.
One out of three hexagons in the planar
unit cell gets ``rectangularly'' deformed along its Sn--Se1
main diagonal; the neighbors on both sides of this plane
adjust to this movement, and two other hexagons suffer
a complicated deformation. Since the main Sn--Se1 axes
in two adjacent planes stand at (approximately) 120$^{\circ}$
to each other, the above mirror symmetry in each layer is not perfect. 
Moreover, the axial compressions/dilations of the ``symmeric''
hexagon go interchanged layer by layer.   
In total, this vibration mode offers a rich pattern of stretchings and
bendings of cation--anion bonds; moreover, it reveals a clear
combined movement of Se atoms against \emph{both} Cu and Sn.
The resulting vector of the cation--anion separation 
(i.e., the varying net dipole moment) goes along [010].
This mode does not have obvious counterparts in CZTSe-kesterite,
and it seems to be important one for spectroscopic identification
of the CTSe phase, capable to yield a big spectral signal.

Vibrations in the \textbf{mode~\#54} (211.7~cm$^{-1}$)
do also occur essentially in the (001) plane;
the (almost) in-plane atomic displacement patterns are shown
in Fig.~\ref{fig:modes49+54+60}, middle row.
The principal stretching in this mode is that of the Cu1--Se1
bonds, accompanied by an asymmetric distortion of the
neighboring hexagons. When coming from plane to plane, the direction
of this stretching bond turns again (in alternation)
by about 120$^{\circ}$ back and forth. In fact, one notices
the cations-vs.-anions movement along ``heterogeneous'' stripes 
-Cu2--Se3--Cu1--Se1--Cu2- which run in the [100] direction
(e.g., two such chains flank the horizontal edges of the fragment
cut in Fig.~\ref{fig:modes49+54+60} for the upper layer). 
Sn and Se2 atoms sitting on the middle line between such stripes
are not much affected by the vibration, but still, the net movement of Sn
is in the same sense as other cations, that is, along [100],
perpendicularly to the dipole moment variation in mode~\#49.

These two modes seem to be two manifestations of the generic
``all-cations-agains-all-anions'' $\Gamma$-TO mode, triply degenerate
in zincblende structure but having the degeneracy lifted as
the symmetry is reduced. In the kesterite phase, the related mode
was \#18 at 216~cm$^{-1}$, singled out because of tetragonal
symmetry. 

The \textbf{mode~\#60} (219.4~cm$^{-1}$) does, in a sense, 
``complement'' the mode~\#49 vibration, in that Sn, now
\emph{in phase} with Se1,
moves rouphly \emph{against} the resulting motion of Se2 and Se3,
these two being again, as in the mode~\#49, at right angle
one to the other. Interestingly, one can find similarities
with mode \#20 if reversing the movement of all cations
(or all anions): whereas there we saw a ``rotation'' of hexagons
due to neighboring cations and anions moving in phase, 
now we find ``benzene-like'' deformation due to anions moving
in opposite to cations and thus stretching the bonds 
in alternation along the hexagons' perimeter.

\begin{figure}
\centerline{\includegraphics[width=0.82\textwidth]{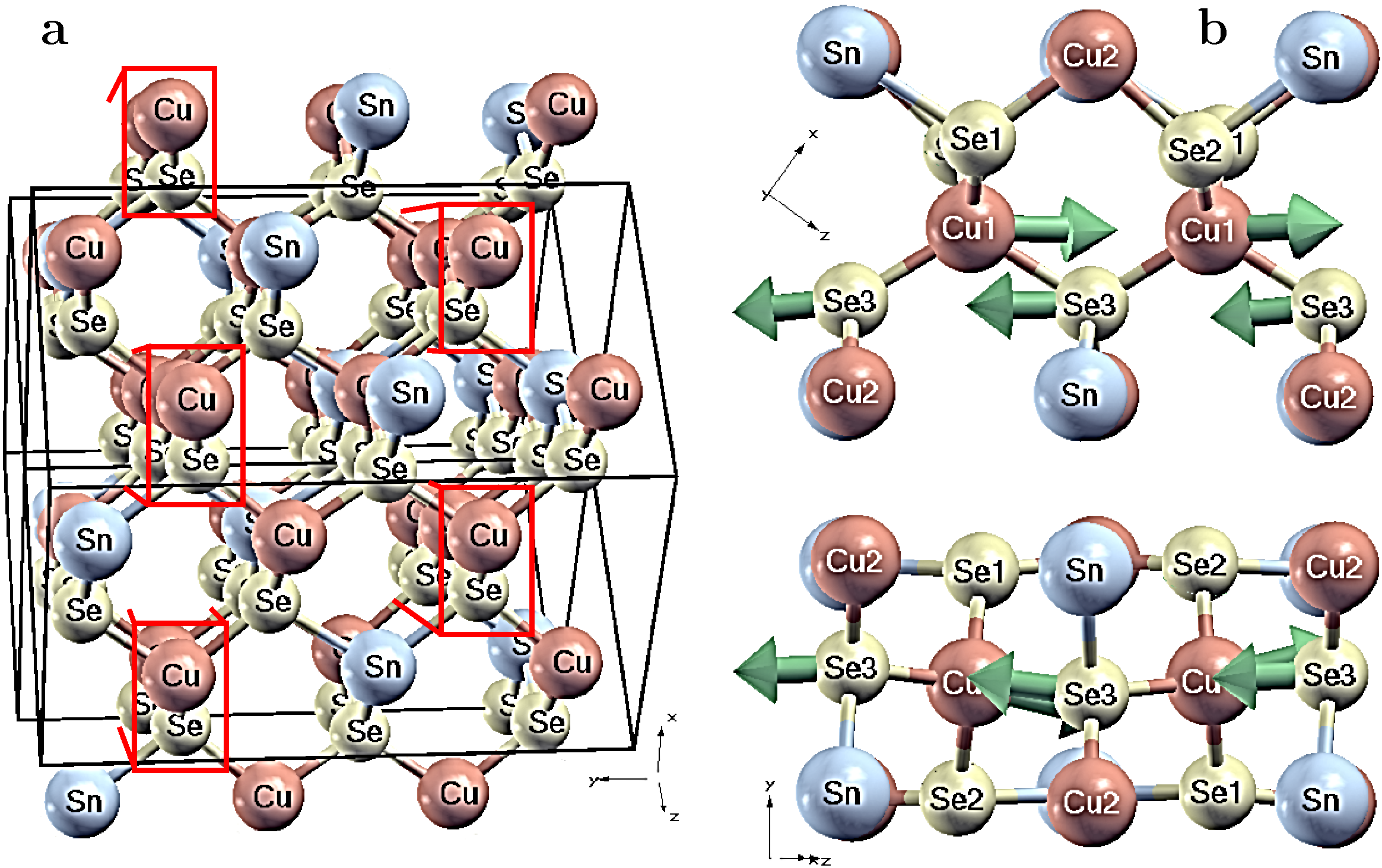}}
\caption{\label{fig:mode66}
(color online)
a) Crystal structure of CTSe, with isolated -Cu1-Se3-Cu1-Se3- chains
running along [101] (perpendicular to the plane of figure)
outlined by red boxes.
b) Snapshot of the vibration mode \#66 at 230~cm$^{-1}$
shown in two projections.}
\end{figure}

We conclude our discussion by 
the \textbf{mode~\#66} (230.2~cm$^{-1}$) which is fully confined 
within the (planar) zigzag -Cu1-Se3-
chains. Fig.~\ref{fig:mode66}a shows how these chains run through
the crystal lattice in the [101] direction. They are separated
from each other by the -Sn-Se1-Cu2-Se2- chains which are out of resonance
at the frequency in question. Fig.~\ref{fig:mode66}b shows the atoms
in the chain with some their neighbors, in the ``side view'' and the
``top view'' (with respect to the orientation of chains
in Fig.~\ref{fig:mode66}a). 
It is well seen, especially in the ``top view'', how the chain
undulates passing through the crystal. An almost degenerate mode 67
(230.5~cm$^{-1}$) differs from the mode 66 in that the movements
within two chains transversing the unit cell occur in counter phase.
We note that such planar -Cu-Se-Cu-Se- chains do not occur in 
the kesterite-type CZTSe, where all the chains are of mixed cation
composition. The closest in nature to the ``along the chain'' vibration
among those in kesterite are the (degenerate) modes 23 and 24 
at 239~cm$^{-1}$ (see Table I and Fig.~6 of Ref.~\cite{PRB82-205204}).
In them, however, the copper is not moving, and hence 
``isolated'' -Se-Zn-Se- fragments vibrate at higher frequency.   
In the stannite-type CZTSe, on the contrary, one finds 
the -Cu-Se-Cu-Se chains
running along the basal diagonals; two other bonds of each Se
are saturated by Zn and Sn in crisscross sequence,
similar to how Cu2 and Sn are attached in Fig.~\ref{fig:mode66}. 
However, no vibration mode of stannite which we've studied 
earlier\cite{PRB82-205204} is any close to a pure along-the-chain
vibration; either the movement of atoms is not quite transversal,
or other out-of-chain atoms are strongly mixed in
(see Table~II and Fig.~7 of Ref.~\cite{PRB82-205204}).
Therefore, the mode \#66 is expected to be a strong signature
of the CTSe phase, not typical to kesterite-type nor to
stannite-type CZTSe. In the absence of more detailed identification
based on the analysis of Raman intensities, we are tempted
to identify this more with the peak at 236~cm$^{-1}$
reported by Altosaar \emph{et al.}\cite{PSSA205-167}

The visualization of chains in Fig.~\ref{fig:mode66} 
helps to explain the lowest mode which yields a non-negligible 
zone-center projection in Fig.~\ref{fig:phdos1}, that is,
\textbf{mode~\#12}, mentioned above to have basically 
a zone-boundary acoustic character. The -Cu1-Se3- chains outlined 
in Fig.~\ref{fig:mode66} make, in mode\#12, rigid entities which move
(along the chain direction) against the rest of crystal 
(interconnected -Sn-Se1-Cu2-Se2- chains). The ``optical'' nature of
such vibration comes from the bending of side bonds to the
-Cu1-Se3- chains. 
\section{Conclusion}
Summarizing, we present comparative results of frozen-phonon calculations
for two chemically and structurally related ordered compounds,
``secondary phase'' Cu$_2$SnSe$_3$ 
and kesterite-type Cu$_2$ZnSnSe$_4$.
There is similarity in positioning of three main groups of spectral lines,
which we identify now, in the order of increased frequency, as
manifestations of $i$) mostly acoustic zone-boundary modes;
$ii$) mostly optical bond-bending modes and $iii$)
genuinely optical bond-stretching modes, understanding all ambiguities
of such attribution in case where all modes exhibit more or less mixed behavior. 
Moreover, the ``master'' TO mode (all cations against all anions)
can be found at about the same frequency.
However, a noticeable mismatch in frequencies and character
of some lines between CZTSe and CZSe can be traced 
to different cation arrangements in these systems.
Unexpectedly, the absence of Zn in one of compounds, 
even if having a limited effect on particular spectral lines, 
does not seem to produce such important
overall differences as does different topology of cation sublattices.
The presence of chains or stripes of different composition and
connectivity, along which different modes come to resonance,
seems to have much larger effect.
Specifically, we suggest that the modes like
\#30 (147~cm$^{-1}$) and \#39 (181~cm$^{-1}$),
which ``live'' on infinite homogenous cation-anion stripes 
not existing in kesterite, or mode \#66 (230~cm$^{-1}$),
confined to Cu-Se chains, another unique feature of CTSe,
can serve as lattice-dynamical fingerprint of this latter
compound, the more so that the latter vibration line 
does apparently appear in measured Raman spectra.

\section*{Acknowledgments}
Cooperation with Susanne Siebentritt was crucial for bringing
the secondary phase into our attention, in the context of related studies
on kesterite-structure photovoltaic materials.
We appreciate enlightening correspondence with J\"uri Krustok
and Maarja Grossberg,
who made us aware of their unpublished results.
Moreover, we are grateful to Michael Yakushev for careful reading
of the manuscript and suggestions.
The calculations have been done using computation resources of the PMMS 
at the Paul Verlaine University -- Metz.


\begin{thebibliography}{10}
\expandafter\ifx\csname url\endcsname\relax
  \def\url#1{\texttt{#1}}\fi
\expandafter\ifx\csname urlprefix\endcsname\relax\def\urlprefix{URL }\fi
\expandafter\ifx\csname href\endcsname\relax
  \def\href#1#2{#2} \def\path#1{#1}\fi

\bibitem{PhTRSA369-1840}
L.~M. Peter, Towards sustainable photovoltaics: the search for new materials,
  Phil. Trans. R. Soc. A 369~(1942) (2011) 1840--1856.
\newblock \href {http://dx.doi.org/10.1098/rsta.2010.0348}
  {\path{doi:10.1098/rsta.2010.0348}}.

\bibitem{AdvMat22-E156}
T.~K. Todorov, K.~B. Reuter, D.~B. Mitzi, High-efficiency solar cell with
  earth-abundant liquid-processed absorber, Advanced Materials 22~(20) (2010)
  E156--E159.
\newblock \href {http://dx.doi.org/10.1002/adma.200904155}
  {\path{doi:10.1002/adma.200904155}}.

\bibitem{APL94-041903}
S.~Chen, X.~G. Gong, A.~Walsh, S.-H. Wei, Crystal and electronic band structure
  of {Cu$_2$ZnSn$X_4$} ($x$ = {S} and {Se}) photovoltaic absorbers:
  First-principles insights, Applied Physics Letters 94~(4) (2009) 041903.
\newblock \href {http://dx.doi.org/10.1063/1.3074499}
  {\path{doi:10.1063/1.3074499}}.

\bibitem{JPD43-215403}
P.~A. Fernandes, P.~M.~P. Salom{\'e}, A.~F. {da}~Cunha, A study of ternary
  {Cu$_2$SnS$_3$} and {Cu$_3$Sn$S_4$} thin films prepared by sulfurizing
  stacked metal precursors, J.Phys.D: Appl. Phys. 43~(21) (2010) 215403.
\newblock \href {http://dx.doi.org/10.1088/0022-3727/43/21/215403}
  {\path{doi:10.1088/0022-3727/43/21/215403}}.

\bibitem{TSF519-7513}
T.~Maeda, S.~Nakamura, T.~Wada, First-principles calculations of vacancy
  formation in {In}-free photovoltaic semiconductor {Cu$_2$ZnSnSe$_4$}, Thin
  Solid Films 519~(21) (2011) 7513--7516.
\newblock \href {http://dx.doi.org/10.1016/j.tsf.2011.01.094}
  {\path{doi:10.1016/j.tsf.2011.01.094}}.

\bibitem{TSF519-7394}
M.~Ganchev, J.~Iljina, L.~Kaupmees, T.~Raadik, O.~Volobujeva, A.~Mere,
  M.~Altosaar, J.~Raudoja, E.~Mellikov, Phase composition of selenized
  {Cu$_2$ZnSnSe$_4$} thin films determined by {X}-ray diffraction and {R}aman
  spectroscopy, Thin Solid Films 519~(21) (2011) 7394--7398.
\newblock \href {http://dx.doi.org/10.1016/j.tsf.2011.01.388}
  {\path{doi:10.1016/j.tsf.2011.01.388}}.

\bibitem{PRB77-125208}
O.~Pag{\`e}s, A.~V. Postnikov, M.~Kassem, A.~Chafi, A.~Nassour, S.~Doyen,
  Unification of the phonon mode behavior in semiconductor alloys: Theory and
  ab initio calculations 77~(12) (2008) 125208.
\newblock \href {http://dx.doi.org/DOI: 10.1103/PhysRevB.77.125208}
  {\path{doi:DOI: 10.1103/PhysRevB.77.125208}}.

\bibitem{PRB71-115206}
A.~V. Postnikov, O.~Pag{\`e}s, J.~Hugel, Lattice dynamics of mixed
  semiconductors {(Be,Zn)Se} from first-principles calculations, Physical
  Review B 71~(11) (2005) 115206 (10~pp).
\newblock \href {http://dx.doi.org/DOI: 10.1103/PhysRevB.71.115206}
  {\path{doi:DOI: 10.1103/PhysRevB.71.115206}}.

\bibitem{JAP90-1847}
G.~Marcano, C.~Rinc{\'o}n, L.~M. {de}~{C}halbaud, D.~B. Bracho,
  G.~S{\'a}nchez~P{\'e}rez, \href{http://dx.doi.org/10.1063/1.1383984}{Crystal
  growth and structure, electrical, and optical characterization of the
  semiconductor {Cu$_2$SnSe$_3$}}, Journal of Applied Physics 90~(4) (2001)
  1847 -- 1853.
\newblock \href {http://dx.doi.org/10.1063/1.1383984}
  {\path{doi:10.1063/1.1383984}}.

\bibitem{MatLet53-151}
G.~Marcano, L.~M. {de}~{C}halbaud, C.~Rinc{\'o}n, G.~S{\'a}nchez~P{\'e}rez,
  \href{http://www.sciencedirect.com/science/article/pii/S0167577X01004669}{Cr%
ystal growth and structure of the semiconductor {Cu$_2$SnSe$_3$}}, Materials
  Letters 53~(3) (2002) 151 -- 154.
\newblock \href {http://dx.doi.org/10.1016/S0167-577X(01)00466-9}
  {\path{doi:10.1016/S0167-577X(01)00466-9}}.

\bibitem{MatResBul38-1949}
G.~E. Delgado, A.~J. Mora, G.~Marcano, C.~Rinc{\'o}n,
  \href{http://www.sciencedirect.com/science/article/pii/S0025540803002642}{Cr%
ystal structure refinement of the semiconducting compound {Cu$_2$SnSe$_3$} from
  {X}-ray powder diffraction data}, Materials Research Bulletin 38~(15) (2003)
  1949 -- 1955.
\newblock \href {http://dx.doi.org/10.1016/j.materresbull.2003.09.017}
  {\path{doi:10.1016/j.materresbull.2003.09.017}}.

\bibitem{siesta}
\href{http://www.icmab.es/siesta/}{Siesta homepage}.
\newline\urlprefix\url{http://www.icmab.es/siesta/}

\bibitem{PRB53-R10441}
P.~Ordej{\'o}n, E.~Artacho, J.~M. Soler,
  \href{http://dx.doi.org/10.1103/PhysRevB.53.R10441}{Self-consistent
  order-{$N$} density-functional calculations for very large systems},
  Phys.~Rev.~B 53~(16) (1996) R10441--R10444.
\newblock \href {http://dx.doi.org/10.1103/PhysRevB.53.R10441}
  {\path{doi:10.1103/PhysRevB.53.R10441}}.

\bibitem{JPCM14-2745}
J.~M. Soler, E.~Artacho, J.~D. Gale, A.~Garc{\'{\i}}a, J.~Junquera,
  P.~Ordej{\'o}n, D.~S{\'a}nchez-Portal,
  \href{http://dx.doi.org/10.1088/0953-8984/14/11/302}{The {SIESTA} method for
  \emph{ab initio} order-{$N$} materials simulation}, J.~Phys.:~Condens.~Matter
  14~(11) (2002) 2745--2779.
\newblock \href {http://dx.doi.org/10.1088/0953-8984/14/11/302}
  {\path{doi:10.1088/0953-8984/14/11/302}}.

\bibitem{PRB43-1993}
N.~Troullier, J.~L. Martins, Efficient pseudopotentials for plane-wave
  calculations, Phys.~Rev.~B 43~(3) (1991) 1993--2006.

\bibitem{PRB82-205204}
N.~B. Mortazavi~Amiri, A.~Postnikov, Electronic structure and lattice dynamics
  in kesterite-type {Cu$_2$ZnSnSe$_4$} from first-principles calculations
  82~(20) (2010) 205204.
\newblock \href {http://dx.doi.org/10.1103/PhysRevB.82.205204}
  {\path{doi:10.1103/PhysRevB.82.205204}}.

\bibitem{PRB84-075213}
Y.-T. Zhai, S.~Chen, J.-H. Yang, H.-J. Xiang, X.-G. Gong, A.~Walsh, J.~Kang,
  S.-H. Wei,
  \href{http://link.aps.org/doi/10.1103/PhysRevB.84.075213}{Structural
  diversity and electronic properties of {Cu$_2$Sn$X_3$} ({$X$}={S}, {Se}): A
  first-principles investigation}, Phys. Rev. B 84~(7) (2011) 075213.
\newblock \href {http://dx.doi.org/10.1103/PhysRevB.84.075213}
  {\path{doi:10.1103/PhysRevB.84.075213}}.

\bibitem{PSSA205-167}
M.~Altosaar, J.~Raudoja, K.~Timmo, M.~Danilson, M.~Grossberg, J.~Krustok,
  E.~Mellikov, {Cu$_2$Zn$_{1-x}$Cd$_x$Sn(Se$_{1-y}$S$_y$)$_4$} solid solutions
  as absorber materials for solar cells, physica status solidi (a) 205~(1)
  (2008) 167--170.
\newblock \href {http://dx.doi.org/10.1002/pssa.200776839}
  {\path{doi:10.1002/pssa.200776839}}.

\bibitem{ThinSF517-2489}
M.~Grossberg, J.~Krustok, K.~Timmo, M.~Altosaar,
  \href{http://www.sciencedirect.com/science/article/pii/S0040609008014053}{Ra%
diative recombination in {Cu$_2$ZnSnSe$_4$} monograins studied by
  photoluminiscence spectroscopy}, Thin Solid Films 517~(7) (2009) 2489 --
  2492.
\newblock \href {http://dx.doi.org/10.1016/j.tsf.2008.11.024}
  {\path{doi:10.1016/j.tsf.2008.11.024}}.

\bibitem{SSC151-84}
G.~Marcano, C.~Rinc{\'o}n, S.~A. L{\'o}pez, G.~S{\'a}nchez~P{\'e}rez, J.~L.
  Herrera-P{\'e}rez, J.~G. Mendoza-Alvarez, P.~Rodr{\'{\i}}guez,
  \href{http://www.sciencedirect.com/science/article/pii/S0038109810006058}{{R%
}aman spectrum of monoclinic semiconductor {Cu$_2$SnSe$_3$}}, Solid State
  Communications 151~(1) (2011) 84 -- 86.
\newblock \href {http://dx.doi.org/10.1016/j.ssc.2010.10.015}
  {\path{doi:10.1016/j.ssc.2010.10.015}}.

\bibitem{CanMiner16-131}
S.~R. Hall, J.~T. Szymanski, J.~M. Stewart, Kesterite, {Cu$_2$(Zn,Fe)SnS$_4$},
  and stannite, {Cu$_2$(Fe,Zn)SnS$_4$} , structurally similar but distinct
  minerals, The Canadian Mineralogist 16~(2) (1978) 131--137.

\end{thebibliography}

\appendix
\section[]{Lattice vectors and atom sites of CTS\lowercase{e} and
CZTS\lowercase{e} in different settings}
\begin{figure}
\centerline{\includegraphics[width=0.7\textwidth]{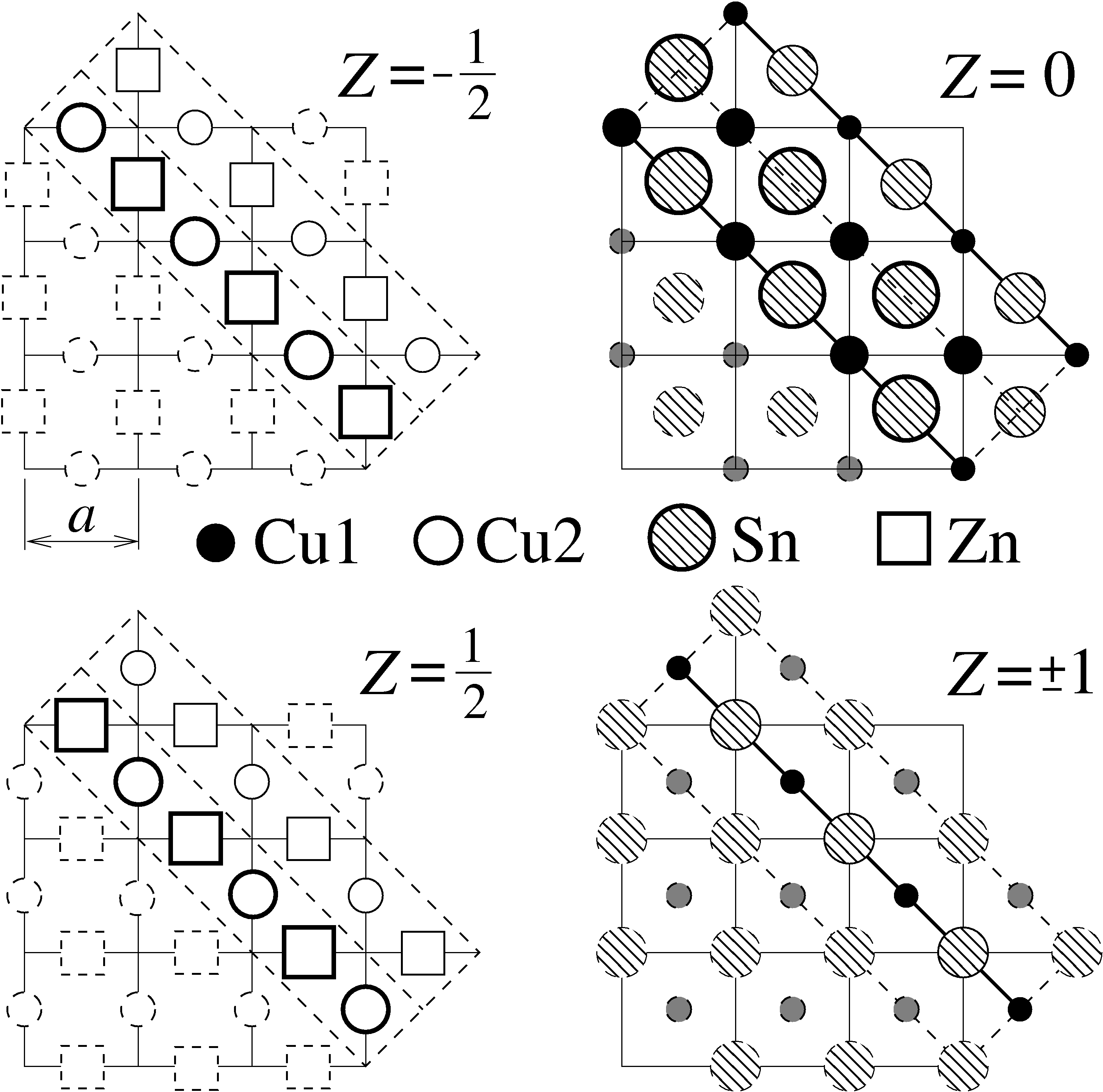}}
\caption{\label{fig:scell1}
Monoclinic unit cell (doubly elongated that of CTSe) sliced by several 
(001) layers of the kesterite structure (only cations are shown). 
Edges of the unit cell which lie in the corresponding planes 
are indicated by thick (diagonal) lines; 
the projection of the rest of the unit cell outline is shown dashed.   
The non-equivalent atoms (listed also in Table~\ref{tab:app2})
in the unit cell are shown large with thick contours; replicated
atoms at the unit cell surface are shown smaller and thinner;
the atoms outside the unit cell have dashed contours. 
}
\end{figure}

Let $\mathbf{a}$, $\mathbf{b}$, $\mathbf{c}$ be lattice vectors of
kesterite structure in conventional setting relative to underlying
cubic (zincblende) lattice, say
$\mathbf{a}=(1\;0\;0)$, $\mathbf{b}=(0\;1\;0)$, 
$\mathbf{c}=(\frac{1}{2}\;\frac{1}{2}\;1)$, 
and cation coordinates as
in Table~\ref{tab:app1}. The monoclinic structure of CTSe is constructed
on translation vectors
$\mathbf{A}=(\frac{1}{2}\;\frac{1}{2}\;1)=\mathbf{c}$,
$\mathbf{C}=(\frac{1}{2}\;\frac{1}{2}\;-1)=
\mathbf{a}+\mathbf{b}-\mathbf{c}$; the third vector we take
doubled, for commensurability with kesterite:
$\mathbf{B}=(3\;-\!3\;\;0) = 3\mathbf{a}-3\mathbf{b}$.
Therefore,
\newcommand{\addspace}{2.5pt} 
\begin{eqnarray}
\left(\begin{array}{c}
\mathbf{A} \\ \mathbf{B} \\ \mathbf{C} \\
\end{array}\right) &=& \left(\begin{array}{rrr}
0 & 0 & 1 \\
3 & -3 & 0 \\
1 & 1 & -1 \\
\end{array}\right)\!
\left(\begin{array}{c}
\mathbf{a} \\ \mathbf{b} \\ \mathbf{c} \\
\end{array}\right); \label{eq:app1}
\\
\left(\begin{array}{c}
\mathbf{a} \\*[{\addspace}] \mathbf{b} \\*[{\addspace}] \mathbf{c} \\
\end{array}\right) &=& \left(\begin{array}{rrr}
\frac{1}{2} &    \frac{1}{6} & \frac{1}{2} \\*[{\addspace}]
\frac{1}{2} & \!-\frac{1}{6} & \frac{1}{2} \\*[{\addspace}]
1 & 0 & 0 \\
\end{array}\right)\!
\left(\begin{array}{c}
\mathbf{A} \\*[{\addspace}] \mathbf{B} \\*[{\addspace}] \mathbf{C} \\
\end{array}\right).
\end{eqnarray}
The monoclinic angle is
$$
\beta=\arccos\frac{2-(c/a)^2}{2+(c/a)^2}\,;
$$
for $c=2a$, $\beta=109.47^{\circ}$,
the tetrahedral bond angle.
The determinant of the transformation matrix in Eq.~(\ref{eq:app1})
is 6, therefore each of the atoms in Table~\ref{tab:app1} has to be
sextupled to find its places over the doubled CTSe cell,
as summarized in Table~\ref{tab:app2}.
Fig.~\ref{fig:scell1} shows the placement of the monoclinic
unit cell over the underlying kesterite structure.

The relation between atom coordinates in Cartesian setting
($X$, $Y$, $Z$) with respect to lattice constant $a$
and fractional ones ($x$, $y$, $z$)
in units of $\mathbf{A}$, $\mathbf{B}$, $\mathbf{C}$
is as follows:
\begin{eqnarray}
\left(\begin{array}{c}
X \\*[{\addspace}] Y \\*[{\addspace}] Z 
\end{array}\right) &=&
\left(\begin{array}{rrr}
\frac{1}{2} &  3 & \frac{1}{2} \\*[{\addspace}]
\frac{1}{2} & -3 & \frac{1}{2} \\*[{\addspace}]
         1  &  0 & -1 
\end{array}\right)\!
\left(\begin{array}{c}
x \\*[{\addspace}] y \\*[{\addspace}] z 
\end{array}\right)\,;
\\
\left(\begin{array}{c}
x \\*[{\addspace}] y \\*[{\addspace}] z 
\end{array}\right) &=&
\left(\begin{array}{rrr}
\frac{1}{2} &  \frac{1}{2} &  \frac{1}{2} \\*[{\addspace}]
\frac{1}{6} & -\frac{1}{6} &  0           \\*[{\addspace}]
\frac{1}{2} &  \frac{1}{2} & -\frac{1}{2}
\end{array}\right)\!
\left(\begin{array}{c}
X \\*[{\addspace}] Y \\*[{\addspace}] Z 
\end{array}\right)\,.
\end{eqnarray}
%


\begin{table}
\caption{\label{tab:app1}
Cation positions in the CZTSe kesterite structure.
Cartesian coordinates are in the units of $a$ (assuming $c=2a$),
fractional coordinates are in the units of $\mathbf{a}$,
$\mathbf{b}$, $\mathbf{c}$.
}
\begin{center}
\begin{tabular}{lcrrrcrrr}
\hline
Atom && \multicolumn{3}{c}{Cartesian} && \multicolumn{3}{c}{Fractional} \\
\hline
Cu1 &&  0 & 0 & 0 && 0 & 0 & 0 \\
Cu2 &&  
0              &  $\frac{1}{2}$ & $\frac{1}{2}$ &&
$-\frac{1}{4}$ &  $\frac{1}{4}$ & $\frac{1}{2}$ \\*[{\addspace}] 
Sn  && 
 $\frac{1}{2}$ &  $\frac{1}{2}$ & 0 &&
 $\frac{1}{2}$ &  $\frac{1}{2}$ & 0 \\*[{\addspace}]
Zn  &&
 $\frac{1}{2}$ &  0             & $\frac{1}{2}$ &&
 $\frac{1}{4}$ & $-\frac{1}{4}$ & $\frac{1}{2}$ \\
\hline
\end{tabular}
\end{center}
\end{table}

\begin{table}
\caption{\label{tab:app2}
Cation coordinates in the CZTSe kesterite structure,
expanded over the doubled mono\-clinic cell.
Cartesian coordinates are given in units of $a/2$,
fractional ones in units of $\mathbf{A}/2$,
$\mathbf{B}/12$, $\mathbf{C}/2$. 
}
\begin{center}
\begin{tabular}{lcrrrcccc}
\hline
Atom && \multicolumn{3}{c}{Cartesian} && \multicolumn{3}{c}{Fractional} \\
\cline{3-6}\cline{7-9}
 &\rule[-2pt]{0mm}{10pt}&
$(2X)$\hspace*{-3mm} & $(2Y)$\hspace*{-3mm} & $(2Z)$\hspace*{-3mm} &&
$(2x)$ & $(12y)$ & $(2z)$ \\
\hline
Cu1 && 0  &    0 &    0 && 0 &  0 & 0  \\
    && 2  &    0 &    0 && 1 &  2 & 1  \\
    && 2  & $-$2 &    0 && 0 &  4 & 0  \\
    && 4  & $-$2 &    0 && 1 &  6 & 1  \\
    && 4  & $-$4 &    0 && 0 &  8 & 0  \\
    && 6  & $-$4 &    0 && 1 & 10 & 1  \\
\hline
Cu2 && 1  &    0 & $-$1 && 0 &  1 & 1  \\
    && 2  & $-$1 &    1 && 1 &  3 & 0  \\
    && 3  & $-$2 & $-$1 && 0 &  5 & 1  \\
    && 4  & $-$3 &    1 && 1 &  7 & 0  \\
    && 5  & $-$4 & $-$1 && 0 &  9 & 1  \\
    && 6  & $-$5 &    1 && 1 & 11 & 0  \\
\hline
Sn  && 1  &    1 &    0 && 1 &  0 & 1  \\
    && 1  & $-$1 &    0 && 0 &  2 & 0  \\
    && 3  & $-$1 &    0 && 1 &  4 & 1  \\
    && 3  & $-$3 &    0 && 0 &  6 & 0  \\
    && 5  & $-$3 &    0 && 1 &  8 & 1  \\
    && 5  & $-$5 &    0 && 0 & 10 & 0  \\
\hline
Zn  && 1  &    0 &    1 && 1 &  1 & 0  \\
    && 2  & $-$1 & $-$1 && 0 &  3 & 1  \\
    && 3  & $-$2 &    1 && 1 &  5 & 0  \\
    && 4  & $-$3 & $-$1 && 0 &  7 & 1  \\
    && 5  & $-$4 &    1 && 1 &  9 & 0  \\
    && 6  & $-$5 & $-$1 && 0 & 11 & 1  \\
    \hline
\end{tabular}
\end{center}
\end{table}

The atomic coordinates of CTSe given in Table~3 of
Delgado \emph{et al.} \cite{MatResBul38-1949} 
look very different from those
in Table~\ref{tab:app2} -- all cation coordinates are far from zero.
In fact, the underlying zincblende structure (if no distinction between 
cations is done), according to Ref.~\cite{MatResBul38-1949}, 
is only slightly distorted (within 2\%), but a rigid shift is applied 
to all coordinates, to satisfy the structure requirements 
of the space group in question. 
The ``nominal'' (undistorted) coordinates in Table~3
of Delgado \emph{et al.} would read as follows:
Cu1 $(\frac{3}{8}\,\frac{1}{4}\,\frac{5}{8})$;
Cu2 $(\frac{3}{8}\,\frac{5}{12}\,\frac{1}{8})$;
Sn $(\frac{3}{8}\,\frac{1}{12}\,\frac{1}{8})$;
Se1 $(0\,\frac{5}{12}\,0)$;
Se2 $(0\,\frac{1}{12}\,0)$;
Se3 $(\frac{1}{2}\,\frac{1}{4}\,0)$. 
We recall that these coordinates are relative to ``short'' (not doubled)
monoclinic cell, with $\mathbf{B}'=(\frac{3}{2}\,-\!\frac{3}{2}\;0)$,
and each site, being of multiplicity 4 as is the only possible
in the space group $Cc$ (Nr.~9), expands into four as follows:
$(x,y,z)$; 
$(\frac{1}{2}\!+\!x, \frac{1}{2}\!+\!y, z)$;
$(x,\bar{y},\frac{1}{2}\!+\!z)$; 
$(\frac{1}{2}\!+\!x,\frac{1}{2}\!-\!y,\frac{1}{2}\!+\!z)$.
 
In order to identify the rigid shift and relate the atoms 
in two different settings, we compare the zincblende 
atomic positions of cations and anions, without discriminating within each
of these groups, according to ``our'' direct counting and to
the above positions, expanded according their multiplicities
in the $Cc$ space group.

\begin{table}
\caption{\label{tab:app3}
Atomic coordinates in zincblende structure,
expanded over the single mono\-clinic cell,
in units of $\mathbf{A}/8$, $\mathbf{B}'/12$, $\mathbf{C}/8$.
The last column indicates the equivalent atom from Table~\ref{tab:app4}
obtained by applying translation $(-\frac{1}{8}\;\frac{1}{4}\;\frac{1}{8})$. 
}
\begin{center}
\begin{tabular}{ccccccc}
\hline
 Atom && \multicolumn{3}{c}{Coordinates} && Equiv. atom \\
 Nr. && $(8x)$ & $(12y)$ & $(8z)$ && in Table~\ref{tab:app4} \\
\hline
\multicolumn{7}{c}{cations} \\
  1 &&  0 &  0 & 0 &&    iv \\
  2 &&  4 &  0 & 4 &&     i \\
  3 &&  4 &  2 & 0 &&     v \\
  4 &&  0 &  2 & 4 &&   xii \\
  5 &&  0 &  4 & 0 &&     x \\
  6 &&  4 &  4 & 4 &&   vii \\
  7 &&  4 &  6 & 0 &&   iii \\
  8 &&  0 &  6 & 4 &&    ii \\
  9 &&  0 &  8 & 0 &&    vi \\
 10 &&  4 &  8 & 4 &&    xi \\
 11 &&  4 & 10 & 0 &&    ix \\
 12 &&  0 & 10 & 4 &&  viii \\
\hline
\multicolumn{7}{c}{anions} \\
 13 &&  1 &  0 & 3 &&  xxiv \\
 14 &&  5 &  0 & 7 &&   xxi \\
 15 &&  1 &  2 & 7 &&  xiii \\
 16 &&  5 &  2 & 3 &&    xx \\
 17 &&  1 &  4 & 3 &&    xv \\
 18 &&  5 &  4 & 7 && xviii \\
 19 &&  1 &  6 & 7 &&  xxii \\
 20 &&  5 &  6 & 3 && xxiii \\
 21 &&  1 &  8 & 3 &&   xix \\
 22 &&  5 &  8 & 7 &&   xiv \\
 23 &&  1 & 10 & 7 &&  xvii \\
 24 &&  5 & 10 & 3 &&   xvi \\ 
\hline
\end{tabular}
\end{center}
\end{table}

\begin{table}
\caption{\label{tab:app4}
Nominal (undistorted) atomic coordinates of monoclinic CTSe,
in units of $\mathbf{A}/8$, $\mathbf{B}'/12$, $\mathbf{C}/8$.
The last column indicates the equivalent atom from Table~\ref{tab:app3}
obtained by translation $(\frac{1}{8}\;\frac{1}{12}\;-\!\frac{1}{8})$. 
}
\begin{center}
\begin{tabular}{lccccccc}
\hline
 & Atom && \multicolumn{3}{c}{Coordinates} && Equiv. atom \\
Type & Nr. && $(8x)$ & $(12y)$ & $(8z)$ && in Table~\ref{tab:app3} \\
\hline
\multicolumn{8}{c}{cations} \\
Cu1 &     i &&  3 &  3 & 5 &&  6 \\
    &    ii &&  7 &  9 & 5 && 12 \\
    &   iii &&  3 &  9 & 1 && 11 \\
    &    iv &&  7 &  3 & 1 &&  5 \\
Cu2 &     v &&  3 &  5 & 1 &&  7 \\
    &    vi &&  7 & 11 & 1 &&  1 \\
    &   vii &&  3 &  7 & 5 && 10 \\
    &  viii &&  7 &  1 & 5 &&  4 \\
Sn  &    ix &&  3 &  1 & 1 &&  3 \\
    &     x &&  7 &  7 & 1 &&  9 \\
    &    xi &&  3 & 11 & 5 &&  2 \\
    &   xii &&  7 &  5 & 5 &&  8 \\
\hline
\multicolumn{8}{c}{anions} \\
Se1 &  xiii &&  0 &  5 & 0 && 19 \\
    &   xiv &&  4 & 11 & 0 && 14 \\
    &    xv &&  0 &  7 & 4 && 21 \\
    &   xvi &&  4 &  1 & 4 && 16 \\
Se2 &  xvii &&  0 &  1 & 0 && 15 \\
    & xviii &&  4 &  7 & 0 && 22 \\
    &   xix &&  0 & 11 & 4 && 13 \\
    &    xx &&  4 &  5 & 4 && 20 \\
Se3 &   xxi &&  4 &  3 & 0 && 18 \\
    &  xxii &&  0 &  9 & 0 && 23 \\
    & xxiii &&  4 &  9 & 4 && 24 \\
    &  xxiv &&  0 &  3 & 4 && 17 \\
    \hline
\end{tabular}
\end{center}
\end{table}

Table~\ref{tab:app3} gives fractional coordinates, in the units
of $\mathbf{A}$, $\mathbf{B}'$, $\mathbf{C}$, of generic cation-anion
zincblende. This amounts to taking half of positions 
from Table~\ref{tab:app2}, namely those with $y<\frac{1}{2}$, and
doubling their $y$-coordinate; moreover, anion positions are added.
The latter are added as in the ``standard'' definition of kesterite
\cite{CanMiner16-131}
with their coordinates in the setting of Table~\ref{tab:app1}
being close to 
($X{\approx}\frac{3}{4}$, $Y{\approx}\frac{3}{4}$, $Z{\approx}\frac{3}{4}$)
rather than to 
($X{\approx}\frac{3}{4}$, $Y{\approx}\frac{3}{4}$, $Z{\approx}\frac{1}{4}$).
The ``mirror symmetric'' placement of anions according to the latter
choice is equally possible; it will simply result
in interchanged $x$ and $z$ coordinates for each atom.

Table~\ref{tab:app4} makes the explicit expansion of ``nominal''
(undisplaced) coordinates of CTSe over the same unit cell.
It is easy to notice that a uniform shift exists,
in fact multiple choices thereof, that would bring the coordinates
of Table~\ref{tab:app3} into those of Table~\ref{tab:app4}.
An example of such attribution, for two different translations,
is indicated in the last column of each table. 

The ``necessity'' of  the shift is imposed by symmetry operations
of the $Cc$ group which permute four atoms within each species. 
In a calculation
which does not make use of crystal symmetry, as our present one,
this shift has no effect whatsoever. We note that the alternative
``mirror'' choice of anionic positions in kesterite would merely
lead to different choice of the uniform shift vector.

\begin{figure}
\centerline{\includegraphics[width=0.65\textwidth]{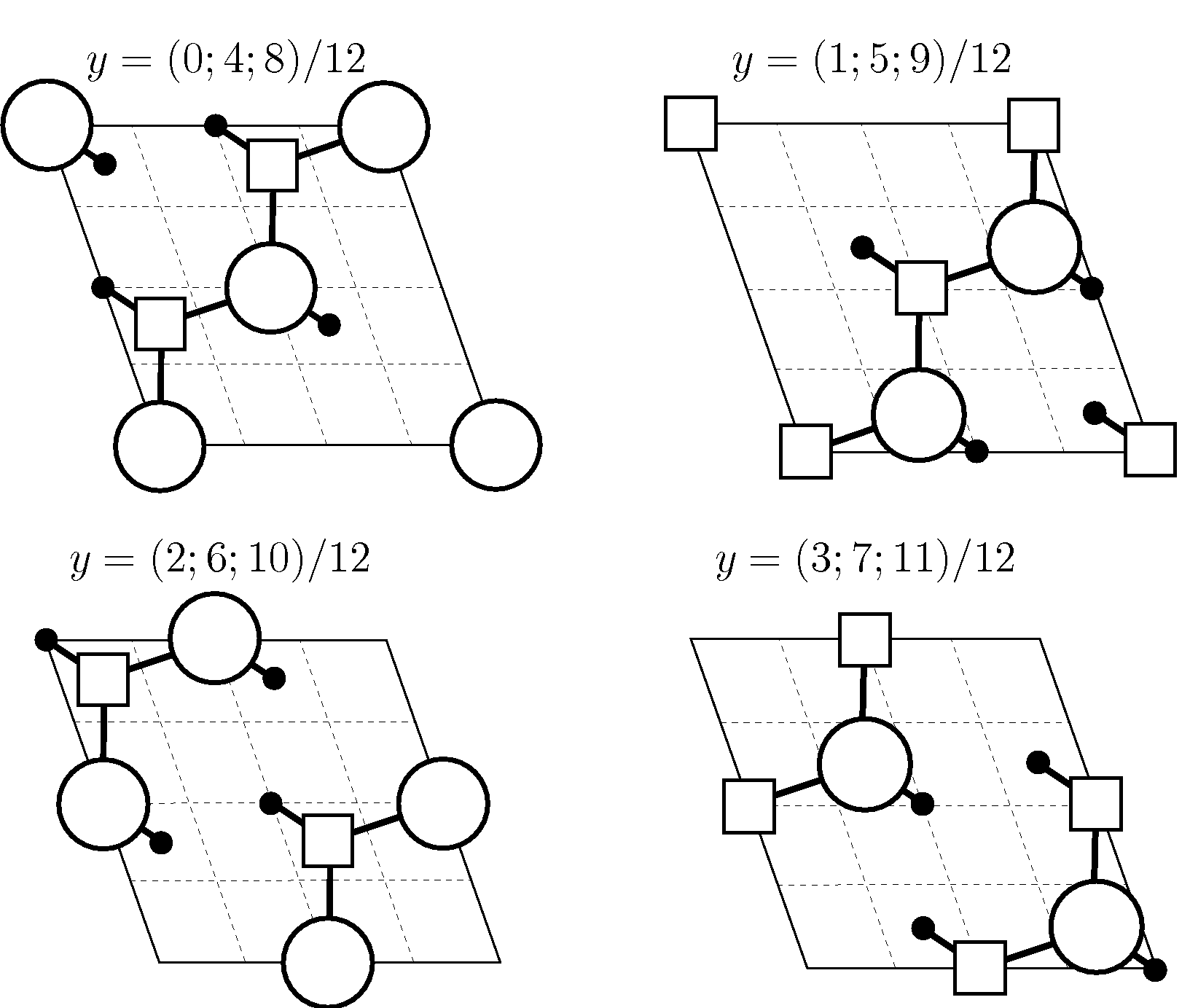}}
\caption{\label{fig:planes1}
Placement of atoms in consecutive (010) planes of the monoclinic
structure of CTSe. Left column: in the ``zincblende'' setting,
as in Table~\ref{tab:app3}. Right column: as in the ``native CTSe''
setting, as in Table~\ref{tab:app4}. Cations are indicated by 
white circles, anions by squares. Thick points mark $(x,y)$ projections
of atoms situated in the adjacent planes below and above. 
}
\end{figure}

More generally, each of two settings is a stacking of
alternating $(010)$ cation-anion planes, 
as shown in Fig.~\ref{fig:planes1}. Each plane is connected
to adjacent ones, forming zigzag (planar) cation-anion chains
along $[010]$; perpendicular ``flat'' similar chains run
in the planes, along $[101]$. One can see that, other than by
introducing a rigid shift, the two settings are related by 
cation$\leftrightarrow$anion interchange, accompanied by
$x{\leftrightarrow}z$ swap, in one of the settings. 

\begin{figure}
\centerline{\includegraphics[width=0.65\textwidth]{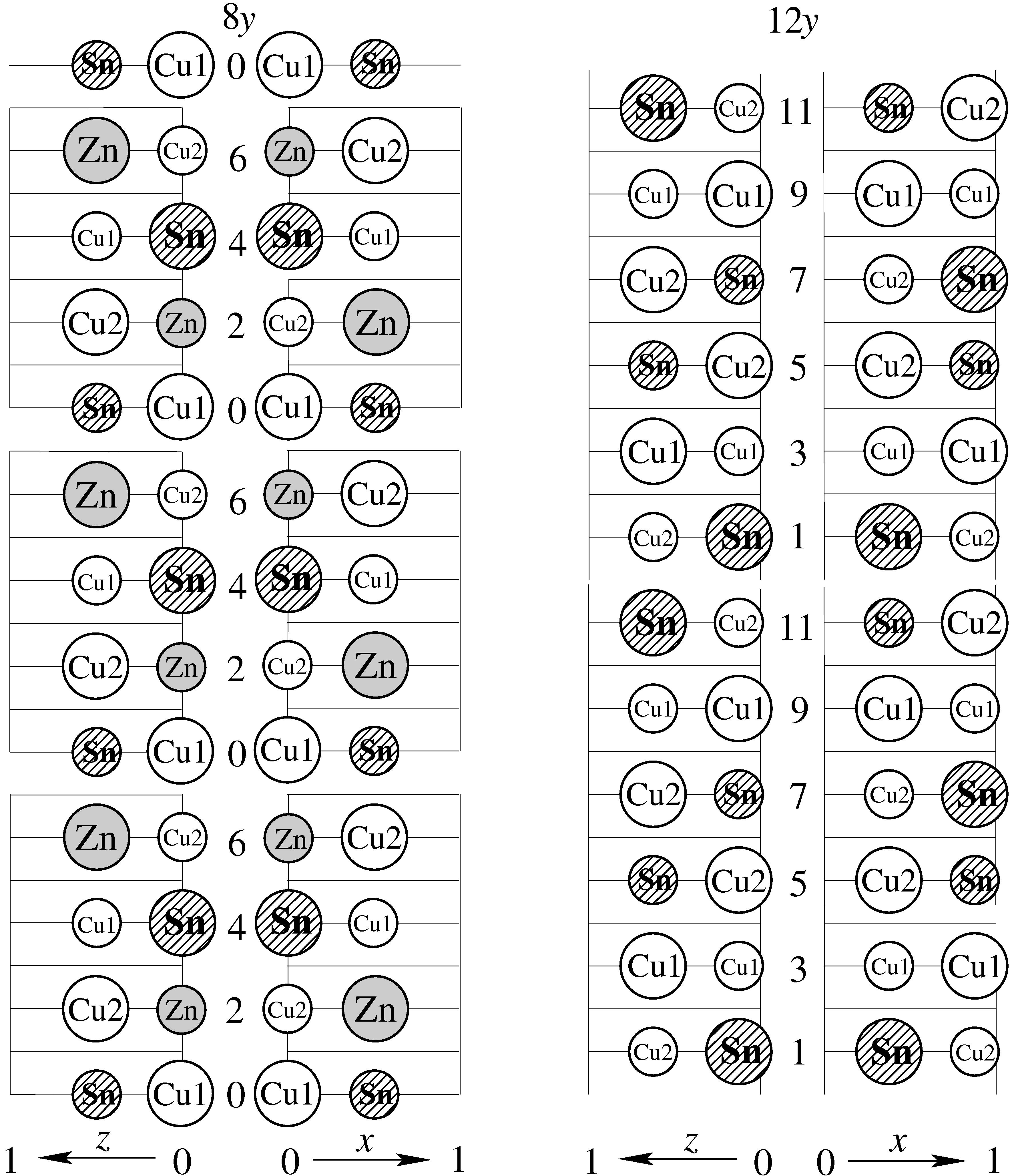}}
\caption{\label{fig:planes2}
Commensurate sequence of $(010)$ planes, in the common
monoclinic setting, of CZTSe kesterite 
(left, three unit cells along $[010]$)
and CTSe (right, two unit cells). Only cations are shown in their
projections on the $(x,y)$ and $(z,y)$ planes. The size of circle
roughly indicates whether the atom is close (large circles) or far
(small circles) from the projection plane. See Fig.~\ref{fig:planes1}
for the $(x,z)$ projections.
}
\end{figure}

We restore now the non-equivalence of cations, recalling that 
they can belong to three chemical species, as in CZTSe kesterite.
A remarkable feature of atom mapping shown by Tables \ref{tab:app3},
\ref{tab:app4}, confirmed also by Fig.~\ref{fig:planes2},
 is that the mapping of a given species 
inevitably occurs into \emph{all} cation species of the counterpart
structure. That is, it is impossible to pinpoint Sn, or Cu,
of CZTSe to their native positions in CTSe and let, say, Zn atom
to appear at whatever cation sites remain vacant. Instead,
the substitution pattern is ondular, along $[010]$: tripled
unit cell of kesterite is commensurate with double unit cell of CTSe.
In the sequence of $(010)$ planes we find Cu-Se ones, common for
both structures; moreover, the CZSe structure contains Cu-Cu planes, 
and kesterite -- Cu-Zn ones. Looking at chains that go along $[010]$
and hold the planes together, we notice that Sn enters only ``isolated''
between two Zn atoms in CZTSe, whereas CTSe contains Sn ``pairs''.
Such similarities and differences find their manifestation 
in the vibration spectra of both systems. 

\end{document}